\documentclass[english,aps,prd,amsmath,amssymb,amsfonts,10pt,onecolumn,a4paper,notitlepage]{revtex4}
\usepackage[T1]{fontenc}
\usepackage[latin1]{inputenc}
\usepackage{bm}
\usepackage{amsmath}
\usepackage{amssymb}
\usepackage{graphicx}
\usepackage{esint}

\makeatletter
\@ifundefined{textcolor}{}
{%
 \definecolor{BLACK}{gray}{0}
 \definecolor{WHITE}{gray}{1}
 \definecolor{RED}{rgb}{1,0,0}
 \definecolor{GREEN}{rgb}{0,1,0}
 \definecolor{BLUE}{rgb}{0,0,1}
 \definecolor{CYAN}{cmyk}{1,0,0,0}
 \definecolor{MAGENTA}{cmyk}{0,1,0,0}
 \definecolor{YELLOW}{cmyk}{0,0,1,0}
 }

\usepackage{amsthm}\usepackage{bm}

\@ifundefined{textcolor}{}{%
 \definecolor{BLACK}{gray}{0}
 \definecolor{WHITE}{gray}{1}
 \definecolor{RED}{rgb}{1,0,0}
 \definecolor{GREEN}{rgb}{0,1,0}
 \definecolor{BLUE}{rgb}{0,0,1}
 \definecolor{CYAN}{cmyk}{1,0,0,0}
 \definecolor{MAGENTA}{cmyk}{0,1,0,0}
 \definecolor{YELLOW}{cmyk}{0,0,1,0}
 }

\def\Mpl{M_{\rm pl}}

\usepackage{babel}

\makeatother

\usepackage{babel}
\begin{document}

\title{Inflationary non-Gaussianities in the most general 
second-order scalar-tensor theories}

\author{Antonio \surname{De Felice}}

\affiliation{Department of Physics, Faculty of Science, Tokyo University of Science, 1-3, 
Kagurazaka, Shinjuku-ku, Tokyo 162-8601, Japan}

\author{Shinji \surname{Tsujikawa}}

\affiliation{Department of Physics, Faculty of Science, Tokyo University of Science, 1-3, 
Kagurazaka, Shinjuku-ku, Tokyo 162-8601, Japan}

\date{\today}

\begin{abstract}

For very general scalar-field theories in which the equations 
of motion are at second-order, we evaluate the 
three-point correlation function of 
primordial scalar perturbations generated during inflation. 
We show that the shape of non-Gaussianities is well approximated 
by the equilateral type. The equilateral non-linear parameter
$f_{\rm NL}^{\rm equil}$ is derived on the quasi de Sitter background
where the slow-variation parameters are much smaller than unity.
We apply our formula for $f_{\rm NL}^{\rm equil}$ to a number of 
single-field models of inflation--such as k-inflation,
k-inflation with Galileon terms, potential-driven Galileon inflation, nonminimal coupling models (including field-derivative coupling models), and Gauss-Bonnet gravity.

\end{abstract}

\maketitle

%%%%%%%%%%%%%%%
\section{Introduction}
%%%%%%%%%%%%%%%

The inflationary paradigm \cite{inflation} can successfully account 
for the observed temperature anisotropies in Cosmic Microwave 
Background (CMB) \cite{WMAP7} as well as the galaxy power spectrum
\cite{LSS}. Although we have not identified the origin 
of inflation yet, the upcoming observations such as PLANCK \cite{PLANCK}
are expected to provide more high-precision data for the 
primordial scalar/tensor power spectra. 
This will allow us to distinguish between a host of theoretical 
models of inflation \cite{inflationmodels} observationally.

In addition to the spectral index of scalar density perturbations 
$n_{\cal R}$ and the tensor-to-scalar ratio $r$, 
the information of primordial scalar non-Gaussianities 
is useful to break the degeneracy 
between different models \cite{Salopek,Gangui,Verde,KSpergel}.
In standard slow-roll inflation in which cosmic 
acceleration is driven by the potential energy 
of a scalar field $\phi$, the non-linear parameter 
$f_{\rm NL}^{\rm equil}$ characterizing the equilateral 
shape of non-Gaussianities is small 
($|f_{\rm NL}^{\rm equil}| \ll 1$) \cite{Bartolo,Maldacena}
(see also Refs.~\cite{Rigo,Lyth,Cremi}).
Meanwhile, in the presence of a non-linear field kinetic energy in 
$X=-\partial^{\mu} \phi \partial_{\mu} \phi/2$ such as 
k-inflation \cite{kinflation}, Dirac-Born-Infeld (DBI) 
models \cite{DBI,DBI2}, (dilatonic) 
ghost condensate \cite{ghost,ghost2}, and Galileon 
inflation \cite{Galileoninf,Galileoninf2},
it is possible to realize large non-Gaussianities 
detectable in future observations
($|f_{\rm NL}^{\rm equil}| \gtrsim 10$) \cite{Seery,Gruzinov,Chen,Ohashi}.

The viable theoretical models of inflation are usually constructed
to keep the equations of motion at second-order in order to avoid 
the Ostrogradski's instability \cite{Ost}. 
In 1974 Horndeski \cite{Horndeski} 
derived the most general single-field Lagrangian giving 
rise to second-order field equations. 
Recently there has been renewed interest in 
second-order gravitational theories in connection 
with the Dvali-Gabadadze-Porrati braneworld \cite{DGP} and 
Galileon gravity \cite{Nicolis,CGalileon} 
(see also Refs.~\cite{Rham,genesis,Kamada,Naruko,Van,Trodden,petel,Elliston,Gao,Galileondark}).
Deffayet {\it et al.} \cite{DGSZ} showed that the most general 
action in those theories is given by 
\begin{equation}
S = \int d^{4}x\sqrt{-g} \biggl[\frac{\Mpl^{2}}{2}\, R
+P(\phi,X)-G_{3}(\phi,X)\,\Box\phi 
+\mathcal{L}_{4}+\mathcal{L}_{5} \biggr].
\label{action}
\end{equation}
Here $g$ is the determinant of the metric $g_{\mu \nu}$, 
$M_{\rm pl}$ is the reduced Planck mass, $R$ is a 
Ricci scalar, and 
\begin{eqnarray}
\mathcal{L}_{4} & = & G_{4}(\phi,X)\, R+G_{4,X}\,[(\Box\phi)^{2}
-(\nabla_{\mu}\nabla_{\nu}\phi)\,(\nabla^{\mu}\nabla^{\nu}\phi)]\,,\\
\mathcal{L}_{5} & = & G_{5}(\phi,X)\, G_{\mu\nu}\,(\nabla^{\mu}\nabla^{\nu}\phi)
-\frac{1}{6}G_{5,X}[(\Box\phi)^{3} 
-3(\Box\phi)\,(\nabla_{\mu}\nabla_{\nu}\phi)\,(\nabla^{\mu}\nabla^{\nu}\phi)
+2(\nabla^{\mu}\nabla_{\alpha}\phi)\,(\nabla^{\alpha}\nabla_{\beta}\phi)\,
(\nabla^{\beta}\nabla_{\mu}\phi)],
\end{eqnarray}
where $G_i$ ($i=3, 4, 5$) are functions in terms of $\phi$ and 
$X=-\partial^{\mu} \phi \partial_{\mu} \phi/2$ with the partial 
derivatives $G_{i,X} \equiv \partial G_i/\partial X$, 
and $G_{\mu \nu}=R_{\mu \nu}-g_{\mu \nu}R/2$ is the Einstein tensor
($R_{\mu \nu}$ is the Ricci tensor).
The action (\ref{action}) is equivalent to that 
derived by Hordenski \cite{Char,KYY}.
In Ref.~\cite{KYY} the spectra of scalar and tensor perturbations 
were derived for the above models.

In this paper we compute the three-point correlation function 
of scalar metric perturbations for the inflationary models 
described by the action (\ref{action}).
Note that the calculation of 
scalar non-Gaussianities up to the term $G_3 (\phi,X) \square \phi$
was carried out in Refs.~\cite{nongaugali,nongaugali2,nongaugali3}
(see also Ref.~\cite{Qiu} for non-minimal coupling models
with $F(\phi)R$).
Taking into account the Lagrangians ${\cal L}_4$ and 
${\cal L}_5$, we can cover a wide variety of gravitational 
theories such as scalar-tensor theories \cite{BD} ($G_4=F(\phi)$), 
field derivative couplings with 
gravity \cite{Germani,Germani2} ($G_5=G(\phi)$),
Galileon gravity \cite{Nicolis,CGalileon} 
($G_3 \propto X$, $G_4 \propto X^2$,
$G_5 \propto X^2$), higher-curvature gravity \cite{fRreview} (including 
Gauss-Bonnet and $f(R)$ theories), $\alpha'$ corrections 
appearing in low-energy effective string theory \cite{Tseytlin}.

Even in the presence of the new contributions coming from 
the terms $G_4$ and $G_5$ we shall show that the shape 
of non-Gaussianities can be well approximated by 
the equilateral type.
The equilateral non-linear parameter $f_{\rm NL}^{\rm equil}$ 
is derived on the quasi de Sitter background.
We also apply our formula to concrete inflationary models
in order to present the cases in which large non-Gaussianities 
can be realized.

This paper is organized as follows.
In Sec.~\ref{modelsec} we derive the background equations of motion 
on the flat Friedmann-Lema\^{i}tre-Robertson-Walker
(FLRW) background and introduce a number of ``slow-variation'' 
parameters convenient for the calculation of primordial non-Gaussianities.
In Sec.~\ref{sactionsec} the action (\ref{action}) is 
expanded at second-order
in perturbations in order to obtain the power spectrum of curvature perturbations
generated during inflation. We also obtain the spectrum of 
gravitational waves and the tensor-to-scalar ratio.
In Sec.~\ref{thirdsec} we expand the action (\ref{action}) at third-order
in perturbations and compute the three-point correlation function of 
curvature perturbations.
In Sec.~\ref{shapesec} we study the shape of non-Gaussianities
for the new terms appearing in the three-point correlation function.
In Sec.~\ref{quasisec} we derive an explicit formula for the 
equilateral non-linear parameter $f_{\rm NL}^{\rm equil}$
under the slow-variation approximation.
In Sec.~\ref{applysec} our formula for $f_{\rm NL}^{\rm equil}$
is applied to a number of concrete models of inflation.
Sec.~\ref{concludesec} is devoted to conclusions.

\section{Background equations}
\label{modelsec}

We first derive the background equations for the theories 
given by the action (\ref{action}) on the flat FLRW background
with the scale factor $a(t)$, where $t$ is cosmic time.
Taking variation of the action at first order with respect to 
the metric elements $g_{00}$, $g_{ii}$,
and the field $\phi$, where $g_{00}$ corresponds to 
the Lapse function $N$, we find
\begin{eqnarray}
E_{1} & \equiv & 
3\Mpl^2 H^2 F+P+6\, H\, G_{{4,\phi}}\dot{\phi}+
\left(G_{{3,\phi}}-12\,{H}^{2}G_{{4,X}}+9\,{H}^{2}G_{{5,\phi}}
-P_{{,X}}\right){\dot{\phi}^{2}} \nonumber \\
& &+\left(6\, G_{{4,\phi X}}-3\, G_{{3,X}}-5\, G_{{5,X}}{H}^{2}\right)
H{\dot{\phi}^{3}}
+3\left(G_{{5,\phi X}}-2\, G_{{4,{\it XX}}}\right)H^{2}\dot{\phi}^{4}
-{H}^{3}G_{{5,{\it XX}}} {\dot{\phi}^{5}}=0\,,
\label{E1} \\
E_{2} & \equiv & 2[(G_{{5,\phi}}-2\, G_{{4,X}}){\dot{\phi}^{2}}
-H G_{{5,X}} {\dot{\phi}^{3}}
+F \Mpl^{2}]\dot{H}+3\Mpl^2 H^2 F+P+4H G_{4,\phi} \dot{\phi}
\nonumber \\
&  & {}+[2\, G_{{4,\phi}}+4H(G_{{5,\phi}}-G_{{4,X}})\dot{\phi}
+(2\, G_{{4,\phi X}}-G_{{3,X}}-3\,{H}^{2}G_{{5,X}}){\dot{\phi}^{2}}
+2H(G_{{5,\phi X}}-2\, G_{{4,{\it XX}}}){\dot{\phi}^{3}}
-H^2 G_{5,XX}\dot{\phi}^4]\ddot{\phi}
\nonumber \\
 &  & {}+\left(2\, G_{{4,\phi\phi}}+3\,{H}^{2}G_{{5,\phi}}-G_{{3,\phi}}
 -6\,{H}^{2}G_{{4,X}}\right){\dot{\phi}^{2}}
 +2H\left(G_{{5,\phi\phi}}-G_{{5,X}}{H}^{2}-2\, G_{{4,\phi X}}\right){\dot{\phi}^{3}}
 -{H}^{2} G_{{5,\phi X}} {\dot{\phi}^{4}}=0\,,
 \label{E2} \\
E_{3} & \equiv & 
[6G_{{4,\phi}}+12(G_{{5,\phi}}-G_{{4,X}})H\dot{\phi}+
3(2\, G_{{4,\phi X}}-G_{{3,X}}-3H^{2}G_{{5,X}})\dot{\phi}^{2}
+6(G_{{5,\phi X}}-2\, G_{{4,{\it XX}}})H\dot{\phi}^{3}
-3H^{2} G_{{5,{\it XX}}} \dot{\phi}^{4}
]\dot{H}\nonumber \\
 &  & {}+\{3(G_{{5,\phi{\it XX}}}-2\, G_{{4,{\it XXX}}}){H}^{2}\dot{\phi}^{4}-{H}^{3}G_{{5,{\it XXX}}}\dot{\phi}^{5}+[3(2\, G_{{4,\phi{\it XX}}}-G_{{3,{\it XX}}})-7\,{H}^{2}G_{{5,{\it XX}}}]H\dot{\phi}^{3}+2\, G_{{3,\phi}}-P_{{,X}}\nonumber \\
 &  & {}+[3(5\, G_{{5,\phi X}}-8\, G_{{4,{\it XX}}}){H}^{2}+G_{{3,\phi X}}-P_{{\it ,XX}}]\dot{\phi}^{2}+6(3\, G_{{4,\phi X}}-G_{{5,X}}{H}^{2}-G_{{3,X}})H\dot{\phi}+6(G_{{5,\phi}}-G_{{4,X}}){H}^{2}\}\ddot{\phi}\nonumber \\
 &  & {}+3(G_{{5,\phi\phi X}}-{H}^{2}G_{{5,{\it XX}}}-2\, G_{{4,\phi{\it XX}}})H^{2}\dot{\phi}^{4}+[(7\, G_{{5,\phi X}}-18\, G_{{4,{\it XX}}}){H}^{2}+3(2\, G_{{4,\phi\phi X}}-G_{{3,\phi X}})]H\dot{\phi}^{3}\nonumber \\
 &  & {}+[3(G_{{5,\phi\phi}}+6\, G_{{4,\phi X}}-3\, G_{{3,X}}){H}^{2}-9\,{H}^{4}G_{{5,X}}-P_{,{\phi X}}+G_{{3,\phi\phi}}]\dot{\phi}^{2}\nonumber \\
 &  & {}+3[6(G_{{5,\phi}}-G_{{4,X}}){H}^{2}+2\, G_{{3,\phi}}-P_{{,X}}]H\dot{\phi}-{H}^{3}G_{{5,\phi{\it XX}}}\dot{\phi}^{5}+12\,{H}^{2}G_{{4,\phi}}+P_{{,\phi}}=0\,,
\label{E3}
\end{eqnarray}
where $H=\dot{a}/a$ is the Hubble parameter, and 
\begin{equation}
F=1+\frac{2G_4}{\Mpl^2}\,.
\end{equation}
Note that a dot represents a derivative with respect to $t$, 
whereas a comma corresponds to a derivative in terms of $\phi$ or $X$
(e.g., $G_{5,\phi X} =\partial^2 G_5/\partial X \partial \phi$).
These equations, because of the Bianchi identities, are not independent
as one can directly verify that 
\begin{equation}
\dot{\phi}E_{3}-\dot{E}_{1}-3H(E_{1}-E_{2})=0\,.
\end{equation}

Eliminating the terms $P$ from Eqs.~(\ref{E1}) and (\ref{E2}), 
it follows that
\begin{eqnarray}
(1-4\delta_{G4X}-2\delta_{G5X}+2\delta_{G5\phi})\epsilon & = & 
\delta_{PX}+3\delta_{G3X}-2 \delta_{G3\phi}
+6\,\delta_{G4X}-\delta_{G4\phi}-6\,\delta_{G5\phi}
+3\,\delta_{G5X}+12\,\delta_{G4XX}+2\,\delta_{G5XX} \nonumber \\
& & -10\,\delta_{G4\phi X}+2\,\delta_{G4\phi\phi}
-8\,\delta_{G5\phi X}+2\,\delta_{G5\phi\phi}
-\delta_{\phi}(\delta_{G3X}+4\,\delta_{G4X}-\delta_{G4\phi} \nonumber \\
& & +8\,\delta_{G4XX}+3\,\delta_{G5X}-4\,\delta_{G5\phi}
+2\,\delta_{G5XX}-2\delta_{G4\phi X}-4\,\delta_{G5\phi X})\,,
\label{epsilonre}
\end{eqnarray}
where we have defined the slow-variation parameters
\begin{eqnarray}
\hspace{-0.4cm}&  & \epsilon=-\frac{\dot{H}}{H^{2}}\,,\qquad
\delta_{\phi}=\frac{\ddot{\phi}}{H\dot{\phi}}\,, \qquad
\delta_{PX}=\frac{P_{,X}X}{\Mpl^{2}H^{2}F}\,,\qquad
\delta_{G3X}=\frac{G_{3,X}\dot{\phi}X}{\Mpl^{2}HF}\,,\qquad
\delta_{G3\phi}=\frac{G_{3,\phi}X}{\Mpl^{2}H^{2}F}\,,\qquad 
\delta_{G4X}=\frac{G_{4,X}X}{\Mpl^{2}F}\,, \nonumber \\
\hspace{-0.4cm}& & \delta_{G4\phi}=\frac{G_{4,\phi}\dot{\phi}}{\Mpl^{2}HF}\,,\qquad\delta_{G4\phi X}=\frac{G_{4,\phi X}\dot{\phi}X}{\Mpl^{2}HF}\,,\qquad\delta_{G4\phi\phi}=\frac{G_{4,\phi\phi}X}{\Mpl^{2}H^{2}F}\,,\qquad\delta_{G4XX}=\frac{G_{4,XX}X^{2}}{\Mpl^{2}F}\,,\qquad
\delta_{G5\phi}=\frac{G_{5,\phi}X}{\Mpl^{2}F}\,,\nonumber \\
\hspace{-0.4cm} &  & \delta_{G5X}=\frac{G_{5,X}H\dot{\phi}X}{\Mpl^{2}F}\,,\qquad
\delta_{G5XX}=\frac{G_{5,XX}H\dot{\phi}X^{2}}{\Mpl^{2}F} \qquad
\delta_{G5\phi X}=\frac{G_{5,\phi X}X^{2}}{\Mpl^{2}F}\,,\qquad\delta_{G5\phi\phi}=\frac{G_{5,\phi\phi}\dot{\phi}X}{\Mpl^{2}HF}\,.
\label{slowva}
\end{eqnarray}
Since $\epsilon \ll 1$ during inflation, we require that the slow-variation parameters 
defined above are much smaller than the order of unity.

Taking the time-derivative of the quantity $\delta_{G4X}$, we obtain 
the first-order quantity
\begin{equation}
\eta_{G4X}\equiv\frac{\dot{\delta}_{G4X}}{H\delta_{G4X}}
=\frac{2\delta_{G4XX}\delta_{\phi}}{\delta_{G4X}}
+\frac{\delta_{G4\phi X}}{\delta_{G4X}}+2\delta_{\phi}-\delta_{F}\,,
\end{equation}
where $\delta_{F}\equiv\dot{F}/(HF)$. 
This means that $\delta_{G4\phi X}$
is higher than the first order. Likewise one can show that 
\begin{equation}
\{ \delta_{G3\phi X},\delta_{G3\phi\phi},
\delta_{G4\phi X},\delta_{G4\phi\phi},\delta_{G5\phi X},
\delta_{G5\phi\phi}\}={\cal O}(\epsilon^{2})\,,
\label{secondorder}
\end{equation}
where 
\begin{equation}
\delta_{G3 \phi X}=\frac{G_{3,\phi X}X^2}{\Mpl^2 H^2 F}\,,
\qquad 
\delta_{G3 \phi \phi}=\frac{G_{3,\phi \phi}\dot{\phi}X}
{\Mpl^2 H^3 F}\,.
\end{equation}
Then, at first order, Eq.~(\ref{epsilonre}) reduces to 
\begin{equation}
\epsilon \simeq  
\delta_{PX}+3\delta_{G3X}-2 \delta_{G3\phi}
+6\,\delta_{G4X}-\delta_{G4\phi}-6\,\delta_{G5\phi}
+3\,\delta_{G5X}+12\,\delta_{G4XX}+2\,\delta_{G5XX}\,.
\label{epap}
\end{equation}
It is also convenient to notice the following relation 
\begin{equation}
\delta_{F}=2\delta_{G4\phi}+4\delta_{G4X}\delta_{\phi} \,,
\label{deltaF}
\end{equation}
that we will use hereafter.

\section{Second-order action}
\label{sactionsec}

In order to discuss the primordial non-Gaussianities we need to first 
study linear perturbation theory. We shall use the momentum
and Hamiltonian constraints to integrate out all the auxiliary fields. 
For the calculation of scalar non-Gaussianities it is convenient to choose
the ADM metric \cite{ADM} in the form 
\begin{equation}
ds^{2}=-[(1+\alpha)^{2}-a(t)^{-2}\, e^{-2{\cal R}}\,(\partial\psi)^{2}]\, dt^{2}
+2\partial_{i}\psi\, dt\, dx^{i}+a(t)^{2}e^{2{\cal R}}d\bm{x}^{2}\,,
\label{eq:metric}
\end{equation}
where $\alpha$, $\psi$, and ${\cal R}$ are scalar 
metric perturbations \cite{Maldacena}.
We choose the uniform field gauge, $\delta\phi=0$, 
to fix the time-component 
of a gauge-transformation vector $\xi^{\mu}$.
The spatial part of $\xi^{\mu}$ is fixed by gauging away a perturbation 
$E$ that appears as a form $E_{,ij}$ in the 
metric (\ref{eq:metric}) \cite{cosmoper}.

Expanding the action (\ref{action}) up to second order
in the perturbations, we find the following result 
\begin{equation}
S_{2}=\int dtd^{3}x\, a^{3}\left[
-3w_{1}\dot{{\cal R}}^{2}+\frac{1}{a^{2}}\,(2w_{1}\dot{{\cal R}}-w_{2}\alpha)\partial^{2}\psi
-\frac{2w_{1}}{a^{2}}\,\alpha\partial^{2}{\cal R}
+3w_{2}\alpha\dot{{\cal R}}+\frac{1}{3}\, w_{3}\alpha^{2}
+\frac{w_{4}}{a^{2}}\,(\partial{\cal R})^{2}\right]\,,
\label{saction}
\end{equation}
 where 
\begin{eqnarray}
w_{1} & = & \Mpl^{2}F-4XG_{4,X}-2HX\dot{\phi}G_{5,X}+2XG_{5,\phi}\,,\\
w_{2} & = & 2\Mpl^{2}HF-2X\dot{\phi}G_{3,X}-16H(XG_{4,X}+X^{2}G_{4,XX})+2\dot{\phi}(G_{4,\phi}+2XG_{4,\phi X})\nonumber \\
 &  & {}-2H^{2}\dot{\phi}(5XG_{5,X}+2X^{2}G_{5,XX})+4HX(3G_{5,\phi}+2XG_{5,\phi X})\,,\\
w_{3} & = & -9M_{{\rm pl}}^{2}H^{2}F+3(XP_{,X}+2X^{2}P_{,XX})+18H\dot{\phi}(2XG_{3,X}+X^{2}G_{3,XX})-6X(G_{3,\phi}+XG_{3,\phi X})\nonumber \\
 &  & +18H^{2}(7XG_{4,X}+16X^{2}G_{4,XX}+4X^{3}G_{4,XXX})-18H\dot{\phi}(G_{4,\phi}+5XG_{4,\phi X}+2X^{2}G_{4,\phi XX})\nonumber \\
 &  & {}+6H^{3}\dot{\phi}(15XG_{5,X}+13X^{2}G_{5,XX}+2X^{3}G_{,5XXX})-18H^{2}X(6G_{5,\phi}+9XG_{5,\phi X}+2X^{2}G_{5,\phi XX})\,,\\
w_{4} & = & \Mpl^{2}F-2XG_{5,\phi}-2XG_{5,X}\ddot{\phi}\,.
\end{eqnarray}

Although the coefficients are quite complicated, 
the structure of the action (\ref{saction}) 
is similar to that found in Ref.~\cite{nongaugali2}.
It is straightforward to find the
momentum and Hamiltonian constraints 
\begin{eqnarray}
\alpha & = & L_{1} \dot{{\cal R}}\,,\label{eq:alpha}\\
\frac{1}{a^{2}}\partial^{2}\psi & = & \frac{2w_{3}}{3w_{2}}\,\alpha+3\dot{{\cal R}}-\frac{2w_{1}}{w_{2}}\,\frac{1}{a^{2}}\,\partial^{2}{\cal R}\,,\label{eq:psi}
\end{eqnarray}
where 
\begin{equation}
L_{1}=\frac{2w_{1}}{w_{2}}\,.
\end{equation}
Substituting Eq.~(\ref{eq:alpha}) into Eq.~(\ref{saction}) and 
making integrations by parts, the second-order action 
reduces to the following form
\begin{equation}
S_{2}=\int dtd^{3}x\, a^{3}
Q\left[\dot{{\cal R}}^{2}-\frac{c_{s}^{2}}{a^{2}}\,(\partial{\cal R})^{2}\right]\,,
\end{equation}
where 
\begin{eqnarray}
Q & = & \frac{w_{1}(4w_{1}w_{3}+9w_{2}^{2})}{3w_{2}^{2}}\,,
\label{Qdef} \\
c_{s}^{2} & = & \frac{3(2w_{1}^{2}w_{2}H-w_{2}^{2}w_{4}
+4w_{1}\dot{w}_{1}w_{2}-2w_{1}^{2}\dot{w}_{2})}
{w_{1}(4w_{1}w_{3}+9w_{2}^{2})}\,.
\end{eqnarray}
In order to avoid the appearance of ghosts and Laplacian 
instabilities we require the conditions 
\begin{equation}
Q>0, \qquad c_s^2>0\,.
\end{equation}

For later convenience we introduce the following parameter 
\begin{equation}
\epsilon_{s}\equiv\frac{Qc_{s}^{2}}{M_{{\rm pl}}^{2}F}=
\frac{2w_{1}^{2}w_{2}H-w_{2}^{2}w_{4}+4w_{1}\dot{w}_{1}w_{2}
-2w_{1}^{2}\dot{w}_{2}}{M_{{\rm pl}}^{2}Fw_{2}^{2}}\,.
\end{equation}
Expansion of $\epsilon_{s}$ in terms of the slow-variation 
parameters leads to 
\begin{eqnarray}
\epsilon_{s} &=& \epsilon+\delta_{G3X}+\delta_{G4\phi}+8\delta_{G4XX}
+\delta_{G5X}+2\delta_{G5XX}+{\cal O}(\epsilon^{2})\nonumber \\
&=& \delta_{PX}+4 \delta_{G3X}-2\delta_{G3 \phi}+6\delta_{G4X}
+20\delta_{G4XX}+4\delta_{G5X}+4\delta_{G5XX}-6\delta_{G5\phi}
+{\cal O}(\epsilon^{2})\,, 
\label{eps}
\end{eqnarray}
where we have used Eqs.~(\ref{secondorder}), (\ref{epap}), and (\ref{deltaF}).

The two-point correlation function of curvature perturbations 
${\cal R}$ can be derived by employing the standard method of 
quantizing the fields on a de Sitter
background \cite{cosmoper}. The power spectrum of ${\cal R}$, 
some time after the Hubble radius crossing, 
is given by 
\begin{equation}
{\cal P}_{{\cal R}}=\frac{H^{2}}{8\pi^{2}Qc_{s}^{3}}
=\frac{H^{2}}{8\pi^{2}\Mpl^{2}F\epsilon_{s}c_{s}}\,.
\label{PR}
\end{equation}
Its spectral index, evaluated at $c_sk=aH$ ($k$ is a comoving wave number),
can be derived as follows
\begin{eqnarray}
n_{{\cal R}}-1 &=& \left.\frac{d\ln{\cal P}_{{\cal R}}}{d\ln k}\right|_{c_{s}k=aH} \nonumber \\
&=& -2\epsilon-\delta_{F}-\eta_{s}-s=
-2\epsilon_{s}-\eta_{s}-s+2\delta_{G3X}+16\delta_{G4XX}
+2\delta_{G5X}+4\delta_{G5XX}+{\cal O}(\epsilon^{2})\,,
\end{eqnarray}
where 
\begin{equation}
\eta_s=\frac{\dot{\epsilon}_s}{H\epsilon_s}\,,\qquad
s=\frac{\dot{c}_s}{H c_s}\,.
\end{equation}
Here we have assumed that $c_s$ is a slowly varying function, 
such that $|s| \ll 1$.

Let us now proceed to the power spectrum of the gravitational waves 
for the theories under consideration. 
We study the tensor perturbations 
of the form
\begin{equation}
ds^{2}=-dt^{2}+a^{2}(t)\,(\delta_{ij}+h_{ij}^{TT})\, dx^{i}\, dx^{j}\,,
\end{equation}
where $h_{ij}^{TT}$ is transverse and traceless.
It is known that the $h_{ij}^{TT}$ can be decomposed into two independent polarization
modes, namely $h_{ij}^{TT}=h_{+}e_{ij}^{+}+h_{\times}e_{ij}^{\times}$.
We choose the normalization for the two matrices such that, in Fourier
space, $e_{ij}^{\lambda}(\bm{k})e_{ij}^{\lambda}(-\bm{k})^{*}=2$,
(where $\lambda={+},{\times}$), and 
$e_{ij}^{+}(\bm{k})e_{ij}^{\times}(-\bm{k})^{*}=0$.
In this case the second-order action for the gravitational
waves can be written as
\begin{equation}
S=\sum_{\lambda}\int dtd^{3}x\, a^{3}Q_{T}\left[\dot{h}_{\lambda}^{2}
-\frac{c_{T}^{2}}{a^{2}}\,(\partial h_{\lambda})^{2}\right]\,,
\end{equation}
where
\begin{eqnarray}
Q_{T} & = & \frac{w_{1}}{4}=\frac{1}{4}\,\Mpl^{2}F\,(1-4\delta_{G4X}
-2\delta_{G5X}+2\delta_{G5\phi})\,,\\
c_{T}^{2} & = & \frac{w_{4}}{w_{1}}=1+4\delta_{G4X}
+2\delta_{G5X}-4\delta_{G5\phi}+{\cal O}(\epsilon^{2})\,.
\end{eqnarray}
Provided that $\{|\delta_{G4X}|, |\delta_{G5X}|, |\delta_{G5\phi}| \}
\ll 1$,
one has $Q_T \simeq \Mpl^{2}F/4$ and hence the no-ghost condition 
$Q_T>0$ is satisfied for $F>0$. Since $c_T^2$ is close to 1, there are
no Laplacian instabilities for tensor perturbations.

The spectrum of tensor perturbations is given by 
\begin{equation}
{\cal P}_{T}=\frac{H^{2}}{2\pi^{2}Q_{T}c_{T}^{3}}
\simeq \frac{2H^{2}}{\pi^{2}\Mpl^{2}F}\,,
\end{equation}
together with the spectral index 
\begin{eqnarray}
n_{T}&=&\left.\frac{d\ln{\cal P}_{T}}{d\ln k}\right|_{c_{T}k=aH} \nonumber \\
&=&-2\epsilon-\delta_{F}=-2\epsilon_{s}+2\delta_{G3X}
+16\delta_{G4XX}+2\delta_{G5X}+4\delta_{G5XX}+{\cal O}(\epsilon^{2})\,.
\end{eqnarray}
For those times during inflation when both ${\cal P}_{T}$ and ${\cal P}_{{\cal R}}$
remain approximately constant, the tensor-to-scalar ratio can be estimated as 
\begin{equation}
r=\frac{{\cal P}_{T}}{{\cal P}_{{\cal R}}} \simeq 16c_{s}\epsilon_{s}\,.
\end{equation}
Then we have the consistency relation
\begin{equation}
r \simeq 8c_{s}(-n_{T}+2\delta_{G3X}+16\delta_{G4XX}
+2\delta_{G5X}+4\delta_{G5XX})\,.
\end{equation}
The consistency relation in standard inflation ($r=-8c_s n_T$) is modified
because of the presence of the terms $G_i$ ($i=3,4,5$).

The spectra of scalar and tensor perturbations coincide with 
those derived in Ref.~\cite{KYY}.

\section{Three-point correlation function}
\label{thirdsec}

We proceed to the calculation of the three-point correlation function 
of curvature perturbations for the theories described by the action 
(\ref{action}). In doing so we expand this action
up to third order in perturbations \cite{Maldacena}.
Although the calculation is quite involved, the steps for 
the derivation of the three-point correlation function 
are similar to those given in detail in Ref.~\cite{nongaugali2}.
The third-order action is given by 
\begin{eqnarray}
S_{3} & = & \int dt\, d^{3}x\, a^{3}\,\{a_{1}\,\alpha^{3}+\alpha^{2}\,(a_{2}\,{\cal R}+a_{3}\,\dot{{\cal R}}+a_{4}\,\partial^{2}{\cal R}/a^{2}+a_{5}\partial^{2}\psi/a^{2})\nonumber \\
 &  & {}+\alpha\,[a_{6}\,\partial_{i}{\cal R}\partial_{i}\psi/a^{2}+a_{7}\,\dot{{\cal R}}{\cal R}+a_{8}\,\dot{{\cal R}}\partial^{2}{\cal R}/a^{2}+a_{9}\,(\partial_{i}\partial_{j}\psi\partial_{i}\partial_{j}\psi-\partial^{2}\psi\partial^{2}\psi)/a^{4}\nonumber \\
 &  & {}+a_{10}(\partial_{i}\partial_{j}\psi\partial_{i}\partial_{j}{\cal R}-\partial^{2}\psi\partial^{2}{\cal R})/a^{4}+a_{11}\,{\cal R}\,\partial^{2}\psi/a^{2}+a_{12}\,\dot{{\cal R}}\,\partial^{2}\psi/a^{2}+a_{13}\,{\cal R}\,\partial^{2}{\cal R}/a^{2}+a_{14}\,(\partial{\cal R})^{2}/a^{2}+a_{15}\dot{{\cal R}}^{2}]\nonumber \\
 &  & {}+b_{1}\,\dot{{\cal R}}^{3}+b_{2}\,{\cal R}\,(\partial{\cal R})^{2}/a^{2}+b_{3}\dot{{\cal R}}^{2}\,{\cal R}+c_{1}\,\dot{{\cal R}}\partial_{i}{\cal R}\partial_{i}\psi/a^{2}+c_{2}\dot{{\cal R}}^{2}\partial^{2}\psi/a^{2}+c_{3}\dot{{\cal R}}\,{\cal R}\,\partial^{2}\psi/a^{2}\nonumber \\
 &  & {}+(d_{1}\dot{{\cal R}}+d_{2}{\cal R})\,(\partial_{i}\partial_{j}\psi\partial_{i}\partial_{j}\psi-\partial^{2}\psi\partial^{2}\psi)/a^{4}+d_{3}\partial_{i}{\cal R}\partial_{i}\psi\,\partial^{2}\psi/a^{4}\}\,,
 \label{eq:S3}
\end{eqnarray}
where 
\begin{eqnarray}
a_{1} & = & 3M_{{\rm pl}}^{2}H^{2}F-(XP_{,X}+4X^{2}P_{,XX}+4X^{3}P_{,XXX}/3)-2H\dot{\phi}(10XG_{3,X}+11X^{2}G_{3,XX}+2X^{3}G_{3,XXX})\nonumber \\
 &  & +2X(G_{3,\phi}+7XG_{3,\phi X}/3+2X^{2}G_{3,\phi XX}/3)-2H^{2}(33XG_{4,X}+126X^{2}G_{4,XX}+68X^{3}G_{4,XXX}+8X^{4}G_{4,XXXX})\nonumber \\
 &  & +2H\dot{\phi}(3G_{4,\phi}+27XG_{4,\phi X}+24X^{2}G_{4,\phi XX}+4X^{3}G_{4,\phi XXX})-H^{3}\dot{\phi}(70XG_{5,X}+98X^{2}G_{5,XX}+32X^{3}G_{5,XXX}\nonumber \\
 &  & +8X^{4}G_{5,XXXX}/3)+2H^{2}X(30G_{5,\phi}+75XG_{5,\phi X}+36X^{2}G_{5,\phi XX}+4X^{3}G_{5,\phi XXX})\,,\\
a_{2} & = & w_{3}\,,\\
a_{3} & = & -3a_{5}\nonumber \\
 & = & -6M_{{\rm pl}}^{2}HF+6\dot{\phi}(2XG_{3,X}+X^{2}G_{3,XX})+12H(7XG_{4,X}+16X^{2}G_{4,XX}+4X^{3}G_{4,XXX})\nonumber \\
 &  & -6\dot{\phi}(G_{4,\phi}+5XG_{4,\phi X}+2X^{2}G_{4,\phi XX})+6H^{2}\dot{\phi}(15XG_{5,X}+13X^{2}G_{5,XX}+2X^{3}G_{5,XXX})\nonumber \\
 &  & -12HX(6G_{5,\phi}+9XG_{5,\phi X}+2X^{2}G_{5,\phi XX})\,,\\
a_{4} & = & -4(XG_{4,X}+2X^{2}G_{4,XX})-8H\dot{\phi}(XG_{5,X}+X^{2}G_{5,XX}/2)+4X(G_{5,\phi}+2XG_{5,\phi X})\,,\\
a_{6} & = & -a_{7}/9=a_{11}=-w_{2}\,,\\
a_{8} & = & 2a_{10}=2b_{1}=-2c_{2}=-4d_{1}=4X\dot{\phi}\, G_{5,X}\,,\\
a_{9} & = & a_{12}/4=-a_{15}/6 \nonumber \\
&=& -M_{{\rm pl}}^{2}F/2+4(XG_{4,X}+X^{2}G_{4,XX})+H\dot{\phi}(5XG_{5,X}+2X^{2}G_{5,XX})-X(3G_{5,\phi}+2XG_{5,\phi X})\,,\\
a_{13} & = & 2a_{14}=2b_{3}/9=-c_{1}=-c_{3}=-4d_{2}/3=d_{3}=-2w_{1}\,,\\
b_{2} & = & w_{4}\,.
\end{eqnarray}
 
We use Eq.~(\ref{eq:alpha}) to eliminate the field $\alpha$ from 
the action (\ref{eq:S3}), which gives
\begin{eqnarray}
S_{3} & = & \int dt\, d^{3}x\, a^{3}\{A_{1}\dot{{\cal R}}^{3}+A_{2}\dot{{\cal R}}^{2}\partial^{2}{\cal R}/a^{2}+A_{3}\dot{{\cal R}}^{2}\partial^{2}\psi/a^{2}+A_{4}{\cal R}\dot{{\cal R}}^{2}+(A_{5}\dot{{\cal R}}+A_{6}{\cal R})\,(\partial_{i}\partial_{j}\psi\partial_{i}\partial_{j}\psi-\partial^{2}\psi\partial^{2}\psi)/a^{4}\nonumber \\
 &  & {}+A_{7}\dot{{\cal R}}(\partial_{i}\partial_{j}\psi\partial_{i}\partial_{j}{\cal R}-\partial^{2}\psi\partial^{2}{\cal R})/a^{4}+A_{8}{\cal R}(\partial{\cal R})^{2}/a^{2}+A_{9}\partial_{i}{\cal R}\partial_{i}\psi\,\partial^{2}\psi/a^{4}\}\,,\label{S3first}
\end{eqnarray}
 where 
\begin{align}
A_{1} & =b_{1}+L_{1}a_{15}+L_{1}^{2}a_{3}+L_{1}^{3}a_{1}\,, & A_{2} & =L_{1}\,(L_{1}a_{4}+a_{8})\,, & A_{3} & =c_{2}+L_{1}a_{12}+L_{1}^{2}a_{5}\,,\nonumber \\
A_{4} & =b_{3}+L_{1}a_{7}+L_{1}^{2}a_{2}=3Q\,, & A_{5} & =L_{1}a_{9}+d_{1}\,, & A_{6} & =d_{2}\,,\nonumber \\
A_{7} & =L_{1}\, a_{10}\,, & A_{8} & =b_{2}+a_{13}\dot{L}_{1}/2+L_{1}(\dot{a}_{13}+Ha_{13})/2\,, & A_{9} & =d_{3}\,.
\end{align}
The next step is to eliminate the field $\psi$ by using 
Eq.~(\ref{eq:psi}). Introducing an auxiliary
field ${\cal X}$ satisfying the relation 
\begin{equation}
\psi=-L_{1}{\cal R}+\frac{a^{2}{\cal X}}{w_1}\,,
\label{psieq}
\end{equation}
it follows that $\partial^{2}{\cal X}=Q\,\dot{{\cal R}}$ from 
Eq.~(\ref{eq:psi}).
Plugging Eq.~(\ref{psieq}) into the action (\ref{S3first}),
we obtain
\begin{equation}
S_{3}=\int dt\, d^{3}x\left(a^{3}f_{1}+af_{2}+f_{3}/a\right)\,,
\label{S3simple}
\end{equation}
where 
\begin{eqnarray}
f_{1} & \equiv & \left(A_{1}+A_{3}\frac{Q}{w_{1}}-A_{5}\frac{Q^{2}}{w_{1}^{2}}\right)\dot{{\cal R}}^{3}+\left(A_{4}-A_{6}\frac{Q^{2}}{w_{1}^{2}}\right){\cal R}\dot{{\cal R}}^{2}+A_{9}\frac{Q}{w_{1}^{2}}\dot{{\cal R}}\partial_{i}{\cal R}\partial_{i}{\cal X}\nonumber \\
 &  & {}+\frac{1}{w_{1}^{2}}\left(A_{5}\dot{{\cal R}}+A_{6}{\cal R}\right)(\partial_{i}\partial_{j}{\cal X})(\partial_{i}\partial_{j}{\cal X})\,,\label{eq:f1expr}\\
f_{2} & \equiv & \left(A_{2}-A_{3}L_{1}+A_{5}\frac{2L_{1}Q}{w_{1}}-A_{7}\frac{Q}{w_{1}}\right)\dot{{\cal R}}^{2}\partial^{2}{\cal R}+A_{6}\frac{2L_{1}Q}{w_{1}}{\cal R}\dot{{\cal R}}\partial^{2}{\cal R}+A_{8}{\cal R}(\partial{\cal R})^{2}-A_{9}\frac{L_{1}Q}{w_{1}}\dot{{\cal R}}(\partial{\cal R})^{2}\nonumber \\
 &  & {}+\frac{A_{7}-2A_{5}L_{1}}{w_{1}}\dot{{\cal R}}(\partial_{i}\partial_{j}{\cal R})(\partial_{i}\partial_{j}{\cal X})-\frac{2A_{6}L_{1}}{w_{1}}\,{\cal R}(\partial_{i}\partial_{j}{\cal R})(\partial_{i}\partial_{j}{\cal X})-\frac{A_{9}L_{1}}{w_{1}}\,\partial^{2}{\cal R}\partial_{i}{\cal R}\partial_{i}{\cal X}\,,\label{eq:f2expr}\\
f_{3} & \equiv & \left(A_{5}L_{1}^{2}-A_{7}L_{1}\right)\dot{{\cal R}}\,[(\partial_{i}\partial_{j}{\cal R})(\partial_{i}\partial_{j}{\cal R})-(\partial^{2}{\cal R})^{2}]+A_{6}L_{1}^{2}{\cal R}\,[(\partial_{i}\partial_{j}{\cal R})(\partial_{i}\partial_{j}{\cal R})-(\partial^{2}{\cal R})^{2}]+A_{9}L_{1}^{2}(\partial{\cal R})^{2}\partial^{2}{\cal R}\,.\label{eq:f3expr}
\end{eqnarray}

Along the same lines of Ref.~\cite{Collins,nongaugali2}
we express the action (\ref{S3simple}) in a more simple form
by carrying out many integrations by parts.
Finally we reach the following action 
\begin{equation}
S_{3}=\int dt\,{\cal L}_{3}\,,
\label{Sfinal}
\end{equation}
where
\begin{eqnarray}
{\cal L}_{3} & = & \int d^{3}x\biggl\{ a^{3}{\cal C}_{1}\Mpl^{2}{\cal R}\dot{{\cal R}}^{2}+a\,{\cal C}_{2}\Mpl^{2}{\cal R}(\partial{\cal R})^{2}+a^{3}{\cal C}_{3}\Mpl\dot{{\cal R}}^{3}+a^{3}{\cal C}_{4}\dot{{\cal R}}(\partial_{i}{\cal R})(\partial_{i}{\cal X})+a^{3}({\cal C}_{5}/\Mpl^{2})\partial^{2}{\cal R}(\partial{\cal X})^{2}\nonumber \\
 &  & {}+a{\cal C}_{6}\dot{{\cal R}}^{2}\partial^{2}{\cal R}+{\cal C}_{7}\left[\partial^{2}{\cal R}(\partial{\cal R})^{2}-{\cal R}\partial_{i}\partial_{j}(\partial_{i}{\cal R})(\partial_{j}{\cal R})\right]/a+a({\cal C}_{8}/\Mpl)\left[\partial^{2}{\cal R}\partial_{i}{\cal R}\partial_{i}{\cal X}-{\cal R}\partial_{i}\partial_{j}(\partial_{i}{\cal R})(\partial_{j}{\cal X})\right]\nonumber \\
 &  & {}+{\cal F}_{1}\frac{\delta{\cal L}_{2}}{\delta{\cal R}}\biggr|_{1}\biggr\}\,,\label{L3}
\end{eqnarray}
and the dimensionless coefficients ${\cal C}_{i}$ ($i=1,\cdots,8$) are 
\begin{eqnarray}
\hspace{-0.5cm}{\cal C}_{1} & = & \frac{1}{\Mpl^{2}}\left[3Q+q_{1}(\dot{Q}+3HQ)-Q\dot{q}_{1}\right],\label{C1f}\\
\hspace{-0.5cm}{\cal C}_{2} & = & \frac{1}{\Mpl^{2}}\left[A_{8}+\frac{1}{a}\frac{d}{dt}(aL_{1}Q)\right],\\
\hspace{-0.5cm}{\cal C}_{3} & = & \frac{1}{\Mpl}\left(A_{1}+A_{3}\frac{Q}{w_{1}}-q_{1}Q\right),\\
\hspace{-0.5cm}{\cal C}_{4} & = & \frac{Q}{w_{1}}\left[\frac{1}{w_{1}}(A_{6}+A_{9})-w_{1}\,\frac{d}{dt}\left(\frac{A_{5}}{w_{1}^{2}}\right)+\frac{3HA_{5}}{w_{1}}\right],\\
\hspace{-0.5cm}{\cal C}_{5} & = & \frac{\Mpl^{2}}{2}\left[\frac{A_{6}}{w_{1}^{2}}-\frac{d}{dt}\left(\frac{A_{5}}{w_{1}^{2}}\right)+\frac{3HA_{5}}{w_{1}^{2}}\right],\\
\hspace{-0.5cm}{\cal C}_{6} & = & A_{2}-A_{3}L_{1}\,,\\
\hspace{-0.5cm}{\cal C}_{7} & = & q_{3}-\frac{Qc_{s}^{2}}{2w_{1}}\,
(A_{7}-2A_{5}L_{1})\,,\\
\hspace{-0.5cm}{\cal C}_{8} & = & \Mpl\left(\frac{q_{2}}{2}-\frac{2c_{s}^{2}A_{5}Q}{w_{1}^{2}}\right).\label{C8f}
\end{eqnarray}
The terms $q_1$, $q_2$, and $q_3$ appear during the step
to derive (\ref{Sfinal}) from (\ref{S3simple}), and are 
given by \cite{nongaugali2}
\begin{eqnarray}
q_1 &=& -\frac{L_1}{c_s^2}\,,\\
q_2 &=& -\frac{4A_{6}L_{1}}{w_{1}}-a^{2}\frac{d}{dt}\!
\left(\frac{A_{7}-2A_{5}L_{1}}{a^{2}w_{1}}\right)-\frac{2A_{9}L_{1}}{w_{1}}\,,\\
q_3 &=& A_{6}L_{1}^{2}-\frac{a}{3}\,\frac{d}{dt}\!
\left(\frac{A_{5}L_{1}^{2}-A_{7}L_{1}}{a}\right)
+\frac{2}{3}\, A_{9}L_{1}^{2}\,.
\end{eqnarray}
The last term in Eq.~(\ref{L3}) is the product of the 
following quantities
\begin{equation}
{\cal F}_{1}=\frac{A_{5}}{w_{1}^{2}}\,\{(\partial_{k}{\cal R})(\partial_{k}{\cal X})-\partial^{-2}\partial_{i}\partial_{j}[(\partial_{i}{\cal R})(\partial_{j}{\cal X})]\}+q_{1}{\cal R}\dot{{\cal R}}+\frac{A_{7}-2A_{5}L_{1}}{4w_{1}a^{2}}\{(\partial{\cal R})^{2}-\partial^{-2}\partial_{i}\partial_{j}[(\partial_{i}{\cal R})(\partial_{j}{\cal R})]\}\,,
\end{equation}
and
\begin{equation}
\frac{\delta{\cal L}_{2}}{\delta{\cal R}}\biggr|_{1}\equiv-2\left[\frac{d}{dt}(a^{3}Q\dot{{\cal R}})-aQc_{s}^{2}\partial^{2}{\cal R}\right]\,.
\label{linearR}
\end{equation}
The coefficient ${\cal F}_1$ includes only the time 
and spatial derivatives of ${\cal R}$ and ${\cal X}$.
In the paper of Maldacena \cite{Maldacena} the term ${\cal R}^2$ is 
also present in the expression of ${\cal F}_1$, which 
gives rise to the contribution of the order of the slow-variation 
parameter in the final expression of the non-linear parameter.
After integrations by parts it is possible to move this 
contribution to other terms such as ${\cal C}_1$ \cite{nongaugali2}, 
which we have done in Eq.~(\ref{L3}).
Hence, as in Refs.~\cite{F1}, we can neglect the contribution 
coming from the term (72).

Defining the conformal time as $\tau=\int a^{-1}\,dt$, 
the vacuum expectation value of curvature perturbations
for the three-point operator at $\tau=\tau_{f}$ 
is \cite{Maldacena,Chen,Koyama}
\begin{equation}
\langle{\cal R}({\bm{k}}_{1}){\cal R}({\bm{k}}_{2}){\cal R}({\bm{k}}_{3})
\rangle=-i\int_{\tau_{i}}^{\tau_{f}}d\tau\, a\,\langle0|\,[{\cal R}(\tau_{f},
{\bm{k}}_{1}){\cal R}(\tau_{f},{\bm{k}}_{2}){\cal R}(\tau_{f},{\bm{k}}_{3}),
{\cal H}_{{\rm int}}(\tau)]\,|0\rangle\,,\label{Rvacuum}
\end{equation}
where the interacting Hamiltonian ${\cal H}_{{\rm int}}$
is given by ${\cal H}_{{\rm int}}=-{\cal L}_{3}$.
Note that $\tau_{i}$ corresponds to the initial time at which 
the perturbations are deep inside the Hubble radius. 
One can take $\tau_{i}\to-\infty$ and $\tau_{f}\to0$
because $\tau\simeq-1/(aH)$ during inflation.
When the integral with respect to $\tau$ is carried out,
we treat the terms ${\cal C}_{i}$ $(i=1,\cdots,8)$
as constants because their variations are small relative to the scale factor $a$.
Each contribution of the three-point correlation function
coming from the terms ${\cal C}_i$ ($i=1,\cdots, 8$) 
is \cite{nongaugali2} 
\begin{eqnarray}
& & \langle{\cal R}({\bm{k}}_{1}){\cal R}({\bm{k}}_{2}){\cal R}({\bm{k}}_{3})\rangle^{(1)}=(2\pi)^{3}\delta^{(3)}({\bm{k}}_{1}+{\bm{k}}_{2}+{\bm{k}}_{3})\frac{{\cal C}_{1}\Mpl^{2}H^{4}}{16Q^{3}c_{s}^{6}}\frac{1}{(k_{1}k_{2}k_{3})^{3}}\left(\frac{k_{2}^{2}k_{3}^{2}}{K}+\frac{k_{1}k_{2}^{2}k_{3}^{2}}{K^{2}}+{\rm sym}\right)\,,\label{H1} \\
& & \langle{\cal R}({\bm{k}}_{1}){\cal R}({\bm{k}}_{2}){\cal R}({\bm{k}}_{3})\rangle^{(2)} = (2\pi)^{3}\delta^{(3)}({\bm{k}}_{1}+{\bm{k}}_{2}+{\bm{k}}_{3})\frac{{\cal C}_{2}\Mpl^{2}H^{4}}{16Q^{3}c_{s}^{8}}\frac{1}{(k_{1}k_{2}k_{3})^{3}}\nonumber \\
& &~~~~~~~~~~~~~~~~~~~~~~~~~~~~~~~~~~
\times\left[({\bm{k}}_{1}\cdot{\bm{k}}_{2}+{\bm{k}}_{2}\cdot{\bm{k}}_{3}+{\bm{k}}_{3}\cdot{\bm{k}}_{1})\left(-K+\frac{k_{1}k_{2}+k_{2}k_{3}+k_{3}k_{1}}{K}+\frac{k_{1}k_{2}k_{3}}{K^{2}}\right)\right]\,,\\
& & \langle{\cal R}({\bm{k}}_{1}){\cal R}({\bm{k}}_{2}){\cal R}({\bm{k}}_{3})\rangle^{(3)}=(2\pi)^{3}\delta^{(3)}({\bm{k}}_{1}+{\bm{k}}_{2}+{\bm{k}}_{3})\frac{3{\cal C}_{3}\Mpl H^{5}}{8Q^{3}c_{s}^{6}}\frac{1}{k_{1}k_{2}k_{3}}\frac{1}{K^{3}}\,,
\\
& & \langle{\cal R}({\bm{k}}_{1}){\cal R}({\bm{k}}_{2}){\cal R}({\bm{k}}_{3})\rangle^{(4)}=(2\pi)^{3}\delta^{(3)}({\bm{k}}_{1}+{\bm{k}}_{2}+{\bm{k}}_{3})\frac{{\cal C}_{4}H^{4}}{32Q^{2}c_{s}^{6}}\frac{1}{(k_{1}k_{2}k_{3})^{3}}\left[\frac{({\bm{k}}_{1}\cdot{\bm{k}}_{2})k_{3}^{2}}{K}\left(2+\frac{k_{1}+k_{2}}{K}\right)+{\rm sym}\right]\,,\\
& & \langle{\cal R}({\bm{k}}_{1}){\cal R}({\bm{k}}_{2}){\cal R}({\bm{k}}_{3})\rangle^{(5)}=(2\pi)^{3}\delta^{(3)}({\bm{k}}_{1}+{\bm{k}}_{2}+{\bm{k}}_{3})\frac{{\cal C}_{5}H^{4}}{16Q\Mpl^{2}c_{s}^{6}}\frac{1}{(k_{1}k_{2}k_{3})^{3}}\left[\frac{k_{1}^{2}({\bm{k}}_{2}\cdot{\bm{k}}_{3})}{K}\left(1+\frac{k_{1}}{K}\right)+{\rm sym}\right]\,,\\
& & \langle{\cal R}({\bm{k}}_{1}){\cal R}({\bm{k}}_{2}){\cal R}({\bm{k}}_{3})\rangle^{(6)}=(2\pi)^{3}\delta^{(3)}({\bm{k}}_{1}+{\bm{k}}_{2}+{\bm{k}}_{3})\frac{3{\cal C}_{6}H^{6}}{4Q^{3}c_{s}^{8}}\frac{1}{k_{1}k_{2}k_{3}}\frac{1}{K^{3}}\,,\\
& & \langle{\cal R}({\bm{k}}_{1}){\cal R}({\bm{k}}_{2}){\cal R}({\bm{k}}_{3})\rangle^{(7)} = (2\pi)^{3}\delta^{(3)}({\bm{k}}_{1}+{\bm{k}}_{2}+{\bm{k}}_{3})\frac{{\cal C}_{7}H^{6}}{8Q^{3}c_{s}^{10}}\frac{1}{(k_{1}k_{2}k_{3})^{3}}\frac{1}{K}\left[1+\frac{k_{1}k_{2}+k_{2}k_{3}+k_{3}k_{1}}{K^{2}}+\frac{3k_{1}k_{2}k_{3}}{K^{3}}\right]\nonumber \\
& &~~~~~~~~~~~~~~~~~~~~~~~~~~~~~~~~~~
\times\left[k_{1}^{2}({\bm{k}}_{2}\cdot{\bm{k}}_{3})-({\bm{k}}_{1}\cdot{\bm{k}}_{2})({\bm{k}}_{1}\cdot{\bm{k}}_{3})+{\rm sym}\right]\,,\\
& & \langle{\cal R}({\bm{k}}_{1}){\cal R}({\bm{k}}_{2}){\cal R}({\bm{k}}_{3})\rangle^{(8)} = (2\pi)^{3}\delta^{(3)}({\bm{k}}_{1}+{\bm{k}}_{2}+{\bm{k}}_{3})\frac{{\cal C}_{8}H^{5}}{32Q^{2}\Mpl c_{s}^{8}}\frac{1}{(k_{1}k_{2}k_{3})^{3}}\frac{1}{K}\nonumber \\
& &~~~~~~~~~~~~~~~~~~~~~~~~~~~~~~~
\times \left\{ \left(2+\frac{2k_{1}+k_{2}+k_{3}}{K}+\frac{2k_{1}(k_{2}+k_{3})}{K^{2}}\right)\left[k_{1}^{2}({\bm{k}}_{2}\cdot{\bm{k}}_{3})-({\bm{k}}_{1}\cdot{\bm{k}}_{2})({\bm{k}}_{1}\cdot{\bm{k}}_{3})\right]+{\rm sym}\right\},\label{H8}
\end{eqnarray}
where $K=k_1+k_2+k_3$.
The symbol ``sym'' corresponds to the symmetric terms
with respect to $k_1$, $k_2$, and $k_3$.

We express the three-point correlation function in the form 
\begin{equation}
\langle{\cal R}({\bm{k}}_{1}){\cal R}({\bm{k}}_{2}){\cal R}({\bm{k}}_{3})\rangle=(2\pi)^{3}\delta^{(3)}({\bm{k}}_{1}+{\bm{k}}_{2}+{\bm{k}}_{3})({\cal P}_{{\cal R}})^{2}B_{{\cal R}}(k_{1},k_{2},k_{3})\,,
\end{equation}
where ${\cal P}_{{\cal R}}$ is given by Eq.~(\ref{PR}), and 
\begin{equation}
B_{{\cal R}}(k_{1},k_{2},k_{3})=\frac{(2\pi)^{4}}{\prod_{i=1}^{3}k_{i}^{3}}{\cal A}_{{\cal R}}\,.
\end{equation}
Collecting all the terms in Eqs.~(\ref{H1})-(\ref{H8}) 
we have
\begin{eqnarray}
\hspace{-1cm}{\cal A}_{{\cal R}} & = & \frac{\Mpl^{2}}{Q}\Biggl\{\frac{1}{4}\left(\frac{2}{K}\sum_{i>j}k_{i}^{2}k_{j}^{2}-\frac{1}{K^{2}}\sum_{i\neq j}k_{i}^{2}k_{j}^{3}\right){\cal C}_{1}+\frac{1}{4c_{s}^{2}}\left(\frac{1}{2}\sum_{i}k_{i}^{3}+\frac{2}{K}\sum_{i>j}k_{i}^{2}k_{j}^{2}-\frac{1}{K^{2}}\sum_{i\neq j}k_{i}^{2}k_{j}^{3}\right){\cal C}_{2}\nonumber \\
 &  & ~~~~+\frac{3}{2}\frac{H}{\Mpl}\frac{(k_{1}k_{2}k_{3})^{2}}{K^{3}}{\cal C}_{3}+\frac{1}{8}\frac{Q}{\Mpl^{2}}\left(\sum_{i}k_{i}^{3}-\frac{1}{2}\sum_{i\neq j}k_{i}k_{j}^{2}-\frac{2}{K^{2}}\sum_{i\neq j}k_{i}^{2}k_{j}^{3}\right){\cal C}_{4}\nonumber \\
 &  & ~~~~+\frac{1}{4}\left(\frac{Q}{\Mpl^{2}}\right)^{2}\,\frac{1}{K^{2}}\left[\sum_{i}k_{i}^{5}+\frac{1}{2}\sum_{i\neq j}k_{i}k_{j}^{4}-\frac{3}{2}\sum_{i\neq j}k_{i}^{2}k_{j}^{3}-k_{1}k_{2}k_{3}\sum_{i>j}k_{i}k_{j}\right]{\cal C}_{5}+\frac{3}{c_{s}^{2}}\left(\frac{H}{\Mpl}\right)^{2}\frac{(k_{1}k_{2}k_{3})^{2}}{K^{3}}{\cal C}_{6}\nonumber \\
 &  & ~~~~+\frac{1}{2c_{s}^{4}}\left(\frac{H}{\Mpl}\right)^{2}\frac{1}{K}\left(1+\frac{1}{K^{2}}\,\sum_{i>j}k_{i}k_{j}+\frac{3k_{1}k_{2}k_{3}}{K^{3}}\right)\left[\frac{3}{4}\,\sum_{i}k_{i}^{4}-\frac{3}{2}\sum_{i>j}k_{i}^{2}k_{j}^{2}\right]\,{\cal C}_{7}\nonumber \\
 &  & ~~~~+\frac{1}{8c_{s}^{2}}\frac{H}{\Mpl}\frac{Q}{\Mpl^{2}}\frac{1}{K^{2}}\left[\frac{3}{2}\, k_{1}k_{2}k_{3}\sum_{i}k_{i}^{2}-\frac{5}{2}\, k_{1}k_{2}k_{3}K^{2}-6\sum_{i\neq j}k_{i}^{2}k_{j}^{3}-\sum_{i}k_{i}^{5}+\frac{7}{2}\, K\sum_{i}k_{i}^{4}\right]{\cal C}_{8}\Biggr\}\,.
\label{AcalR}
\end{eqnarray}
We also define the non-linear parameter $f_{{\rm NL}}$, as \cite{WMAP7,Takamizu}
\begin{equation}
f_{{\rm NL}}=\frac{10}{3}\frac{{\cal A}_{{\cal R}}}
{\sum_{i=1}^{3}k_{i}^{3}}\,.
\end{equation}
If the three-point correlation function is described by the equilateral
configuration ($k_{1}=k_{2}=k_{3}$), it follows that  
\begin{eqnarray}
f_{{\rm NL}} & = & \frac{40}{9}\frac{\Mpl^{2}}{Q}\biggl[\frac{1}{12}{\cal C}_{1}+\frac{17}{96c_{s}^{2}}{\cal C}_{2}+\frac{1}{72}\frac{H}{\Mpl}{\cal C}_{3}-\frac{1}{24}\frac{Q}{\Mpl^{2}}{\cal C}_{4}-\frac{1}{24}\left(\frac{Q}{\Mpl^{2}}\right)^{2}{\cal C}_{5}+\frac{1}{36c_{s}^{2}}\left(\frac{H}{\Mpl}\right)^{2}{\cal C}_{6}\nonumber \\
 &  & ~~~~~~~~~~-\frac{13}{96c_{s}^{4}}\left(\frac{H}{\Mpl}\right)^{2}{\cal C}_{7}-\frac{17}{192c_{s}^{2}}\frac{H}{\Mpl}\frac{Q}{\Mpl^{2}}{\cal C}_{8}\biggr]\,.
 \label{fnl}
\end{eqnarray}
This is the same form as that derived in Ref.~\cite{nongaugali2}, 
but the coefficients ${\cal C}_i$ ($i=1, \cdots, 8$) are different.

\section{The shapes of non-Gaussianities}
\label{shapesec}

The shapes of non-Gaussianities have been discussed by a number of
authors (see e.g.,~Refs.~\cite{Chen,Babich}). 
In Ref.~\cite{Chen} Chen {\it et al.} studied the case of k-inflation,
where only the terms proportional to ${\cal C}_{i}$ 
(with $1 \leq i \leq 5$) are present. 
Here we wish to discuss the shapes coming from the remaining terms, 
that is, we focus our attention only on the following terms 
\begin{eqnarray}
{\cal B}_{{\cal R}}^{(6)} & = & 
\frac{1}{\sum_{i=1}^{3}k_{i}^{3}}\,
\frac{3H^{2}}{Qc_{s}^{2}}\,
\frac{(k_{1}k_{2}k_{3})^{2}}{K^{3}}{\cal C}_{6}\,,\\
{\cal B}_{{\cal R}}^{(7)} & = & 
\frac{1}{\sum_{i=1}^{3}k_{i}^{3}}\,
\frac{H^{2}}{2Qc_{s}^{4}}\,\frac{1}{K}\left(1+\frac{1}{K^{2}}\,\sum_{i>j}k_{i}k_{j}+\frac{3k_{1}k_{2}k_{3}}{K^{3}}\right)\left[\frac{3}{4}\,\sum_{i}k_{i}^{4}-\frac{3}{2}\sum_{i>j}k_{i}^{2}k_{j}^{2}\right]\,{\cal C}_{7}\,,\\
{\cal B}_{{\cal R}}^{(8)} & = & 
\frac{1}{\sum_{i=1}^{3}k_{i}^{3}}\,
\frac{1}{8c_{s}^{2}}\frac{H}{\Mpl}\frac{1}{K^{2}}\left[\frac{3}{2}\, k_{1}k_{2}k_{3}\sum_{i}k_{i}^{2}-\frac{5}{2}\, k_{1}k_{2}k_{3}K^{2}-6\sum_{i\neq j}k_{i}^{2}k_{j}^{3}-\sum_{i}k_{i}^{5}+\frac{7}{2}\, K\sum_{i}k_{i}^{4}\right]{\cal C}_{8}\,.
\end{eqnarray}

Although the total bispectrum is the sum of all the previous contributions
(in addition to the standard ones of k-inflation), we study
the momentum dependence of each single contribution 
${\cal B}_{{\cal R}}^{(i)}$
(with $6\leq i\leq8$). In fact, we will check that the non-Gaussianities
associated to each single term ${\cal B}_{{\cal R}}^{(i)}$ can be well 
measured by means of the equilateral estimator defined as 
\begin{equation}
{\cal B}_{{\cal R}}^{{\rm equil}}\equiv
(2\pi)^{4}\left(\frac{9}{10}f_{\rm NL}^{\rm equil}\right)
\left[-\frac{1}{k_{1}^{3}k_{2}^{3}}-\frac{1}{k_{1}^{3}k_{3}^{3}}-\frac{1}{k_{2}^{3}k_{3}^{3}}-\frac{2}{k_{1}^{2}k_{2}^{2}k_{3}^{2}}+\frac{1}{k_{1}k_{2}^{2}k_{3}^{3}}+(5\;{\rm perms.})\right]\,.
\end{equation}

\begin{figure}
\includegraphics[width=8.5cm]{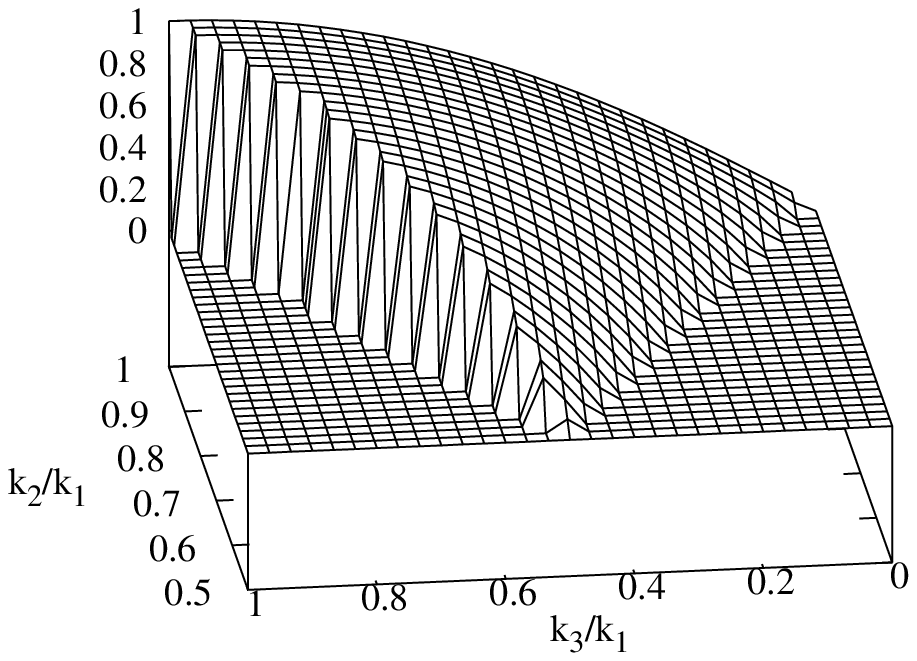} 
\includegraphics[width=8.5cm]{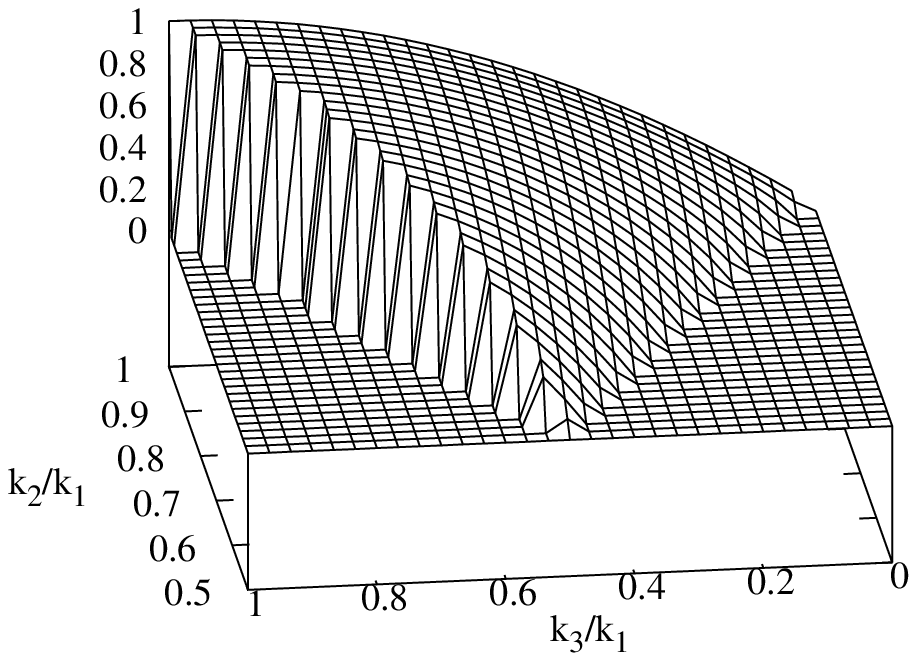}
\caption{The shape functions 
${\cal B}_{{\cal R}}^{(7)}(1,k_{2}/k_{1},k_{3}/k_{1})
(k_{2}/k_{1})^{2}(k_{3}/k_{1})^{2}$ (left) and
${\cal B}_{{\cal R}}^{(8)}(1,k_{2}/k_{1},k_{3}/k_{1})
(k_{2}/k_{1})^{2}(k_{3}/k_{1})^{2}$ (right).
The two functions are plotted in the domain 
$1-k_{2}/k_{1}\leq k_{3}/k_{1}\leq k_{2}/k_{1}$.
The lower boundary is given by the triangular inequality, whereas
the upper boundary is chosen in order not to repeat twice 
the same physical
configuration. Finally, the functions ${\cal B}_{{\cal R}}^{(i)}$
are multiplied by the measure $(k_{2}/k_{1})^{2}(k_{3}/k_{1})^{2}$
following Ref.~\cite{Babich}.
The plots are normalized to have a unit value
at the point $k_{2}/k_{1}=k_{3}/k_{1}=1$.
\label{fig:shapes}}
\end{figure}

In Ref.~\cite{nongaugali} it was shown that ${\cal B}_{{\cal R}}^{(6)}$
exhibits the same momentum dependence as that of the term  
${\cal B}_{{\cal R}}^{(3)}$, which has a high correlation 
with ${\cal B}_{{\cal R}}^{{\rm equil}}$ (about 0.936).
Therefore, we further restrict our analysis only to the new non-trivial
terms ${\cal B}_{{\cal R}}^{(7)}$ and ${\cal B}_{{\cal R}}^{(8)}$.
In Fig.~\ref{fig:shapes} we plot the shapes of 
the bispectrum contributions
${\cal B}_{{\cal R}}^{(7)}$ and ${\cal B}_{{\cal R}}^{(8)}$.

Let us quantify how much the shape functions of 
${\cal B}_{{\cal R}}^{(i)}$
can be fitted with the function ${\cal B}_{{\cal R}}^{{\rm equil}}$.
We follow the procedure shown in Ref.~\cite{Fergusson} and calculate the
shape correlator $C({\cal B}_{{\cal R}},{\cal B}'_{{\cal R}})$ defined as 
\begin{equation}
C({\cal B}_{{\cal R}},{\cal B}'_{{\cal R}})=\frac{{\cal I}
({\cal B}_{{\cal R}},{\cal B}'_{{\cal R}})}
{\sqrt{{\cal I}({\cal B}_{{\cal R}},{\cal B}{}_{{\cal R}})\,
{\cal I}({\cal B}'_{{\cal R}},{\cal B}'_{{\cal R}})}}\,,\label{eq:corr1}
\end{equation}
where 
\begin{equation}
{\cal I}({\cal B}_{{\cal R}},{\cal B}'_{{\cal R}})=
\int d{\cal V}_{k}{\cal B}_{{\cal R}}(k_{1},k_{2},k_{3})\,
{\cal B}'_{{\cal R}}(k_{1},k_{2},k_{3})
\frac{k_{1}^{4}k_{2}^{4}k_{3}^{4}}{(k_{1}+k_{2}+k_{3})^{3}}\,,
\end{equation}
and the region of integration is defined by the condition 
$0\leq k_{1}<\infty$,
$0<k_{2}/k_{1}<1$, and $1-k_{2}/k_{1}\leq k_{3}/k_{1}\leq1$. In
terms of the variables $k_{2}/k_{1}$ and $k_{3}/k_{1}$ the integral
in $k_{1}$ factorizes out.
After performing numerical integrations we find that 
\begin{equation}
C({\cal B}_{{\cal R}}^{(7)},{\cal B}{}_{{\cal R}}^{{\rm equil}})
=-0.99989\,\qquad{\rm and}\qquad C({\cal B}_{{\cal R}}^{(8)},
{\cal B}{}_{{\cal R}}^{{\rm equil}})=-0.99999.
\end{equation}

Since both the contributions ${\cal B}_{{\cal R}}^{(7)}$ and 
${\cal B}_{{\cal R}}^{(8)}$ are almost completely anti-correlated 
with ${\cal B}_{{\cal R}}^{{\rm equil}}$,
it makes sense to use $f_{\rm NL}^{{\rm equil}}$ to measure
the shapes of non-Gaussianities in the presence of the contributions
${\cal B}_{{\cal R}}^{(7)}$ and ${\cal B}_{{\cal R}}^{(8)}$ as well.
We also checked that $C({\cal B}_{{\cal R}}^{(6)},
{\cal B}{}_{{\cal R}}^{{\rm equil}})=0.936177$,
which agrees with the result in Ref.~\cite{nongaugali}.

\section{The non-linear parameter under the slow-variation approximation}
\label{quasisec}

In this section we derive the approximate expression of the equilateral
non-linear parameter $f_{{\rm NL}}^{{\rm equil}}$ with $k_1=k_2=k_3$
on the quasi de Sitter background.
In order to obtain a compact form of $f_{{\rm NL}}^{{\rm equil}}$
we define the following quantities
\begin{eqnarray}
\Sigma &\equiv& \frac{w_1 (4w_1w_3+9w_2^2)}{12\Mpl^4}\,,\\
\lambda &\equiv& \frac{F^2}{3} [ 3X^2 P_{,XX}+2X^3 P_{,XXX}+
3H \dot{\phi} (X G_{3,X}+5X^2 G_{3,XX}+2X^3 G_{3,XXX})
-2(2X^2 G_{3,\phi X}+X^3 G_{3,\phi XX}) \nonumber \\
& &+6H^2 (9X^2 G_{4,XX}+16 X^3G_{4,XXX}+4X^4 G_{4,XXXX})
-3 H\dot{\phi} (3X G_{4 \phi,X}+12 X^2 G_{4,\phi XX}+4X^3 G_{4,\phi XXX})
\nonumber \\
& &+H^3 \dot{\phi} (3X G_{5,X}+27 X^2 G_{5,XX}+24 X^3 G_{5,XXX}
+4X^4 G_{5,XXXX}) \nonumber \\
& &-6 H^2 (6X^2G_{5,\phi X}+9 X^3 G_{5,\phi XX}
+2X^4 G_{5,\phi XXX})]\,,
\end{eqnarray}
which are the generalizations of those introduced
in Refs.~\cite{Seery,nongaugali,nongaugali2}. 
Notice that $\Sigma$ is related with $Q$ via 
$Q=4\Mpl^4 \Sigma/w_2^2$.
We also introduce the following quantities
\begin{equation}
\lambda_{3X}\equiv\frac{XG_{3,XX}}{G_{3,X}}\,,
\qquad\lambda_{4X}\equiv\frac{XG_{4,XXX}}{G_{4,XX}}\,,
\qquad\lambda_{5X}\equiv\frac{XG_{5,XXX}}{G_{5,XX}}\,.
\label{lambdai}
\end{equation}

We derive the leading-order terms to each 
$f_{{\rm NL}}^{{\rm equil}\,(i)}$ coming from the 
coefficients ${\cal C}_i$ ($i=1,\cdots,8$).
We expand each $f_{{\rm NL}}^{{\rm equil}\,(i)}$ 
in terms of the slow-variation parameters defined 
in Eq.~(\ref{slowva}), by treating $c_s^2$,
$\lambda/\Sigma$, and $\lambda_{iX}$ ($i=3,4,5$)
as arbitrary parameters.
Then the leading-order contributions to 
$f_{{\rm NL}}^{{\rm equil}\,(i)}$ are given by 
\begin{eqnarray}
f_{{\rm NL}}^{{\rm equil\,(1)}} & = & \frac{10}{9}\left(1-\frac{1}{c_{s}^{2}}\right)+\frac{10}{27c_{s}^{2}}\left(\epsilon_{s}-\eta_{s}-4\delta_{G3X}-12\delta_{G4X}-32\delta_{G4XX}+12\delta_{G5\phi}-10\delta_{G5X}-8\delta_{G5XX}\right)\,,\\
f_{{\rm NL}}^{{\rm equil\,(2)}} & = & \frac{85}{108}\left(\frac{1}{c_{s}^{2}}-1\right)+\frac{85}{108c_{s}^{2}}\left(\epsilon_{s}+\eta_{s}-2s+4\delta_{G4X}+2\delta_{G5X}-4\delta_{G5\phi}\right)\,,\\
f_{{\rm NL}}^{{\rm equil\,(3)}} & = & \frac{5}{81}\left(\frac{1}{c_{s}^{2}}-1\right)
-\frac{10}{81} \frac{\lambda}{\Sigma}
+\frac{5}{81c_{s}^{2}}(\delta_{G3X}+4\delta_{G4X}+3\delta_{G5X}-\delta_{G4\phi}-4\delta_{G5\phi}+8\delta_{G4XX}+2\delta_{G5XX})
\nonumber \\
 &  & {}-\frac{5}{81}\left(3+2\lambda_{3X}\right)\delta_{G3X}-\frac{40}{81}
 \left(5+2\lambda_{4X}\right)\delta_{G4XX}-\frac{20}{81}(4+\lambda_{5X})
 \delta_{G5XX} \nonumber \\
& & +\frac{5}{81}(\delta_{G4\phi}+8\delta_{G5\phi}-8\delta_{G4X}-9\delta_{G5X})
-\frac{10}{27}\,\frac{c_{s}^{2}}{\epsilon_{s}} \left[
(1+\lambda_{3X})\delta_{G3X}^{2}+\xi (\delta^2) 
\right]\,, \label{eq:fNL3}\\
f_{{\rm NL}}^{{\rm equil\,(4)}} & = & \frac{10}{27}\frac{\epsilon_{s}}{c_{s}^{2}}\,,\\
f_{{\rm NL}}^{{\rm equil\,(5)}} & = & -\frac{5}{108c_{s}^{2}}\left(\epsilon_{s}-4\delta_{G3X}-8\delta_{G4XX}+8\delta_{G5X}+4\delta_{G5XX}\right)\epsilon_{s}\,,\\
f_{{\rm NL}}^{{\rm equil\,(6)}} & = & \frac{20}{81\epsilon_{s}}\left[(1+\lambda_{3X})\delta_{G3X}+4(3+2\lambda_{4X})\delta_{G4XX}+\delta_{G5X}+(5+2\lambda_{5X})\delta_{G5XX}\right]\,,\\
f_{{\rm NL}}^{{\rm equil\,(7)}} & = & \frac{65}{162c_{s}^{2}\epsilon_{s}}\left(\delta_{G3X}+6\delta_{G4XX}+\delta_{G5X}+\delta_{G5XX}\right)\,,\\
f_{{\rm NL}}^{{\rm equil\,(8)}} & = & -\frac{85}{108c_{s}^{2}}\left(\delta_{G3X}+4\delta_{G4XX}\right)\,.
\end{eqnarray}
The explicit expression of the second-order term $\xi (\delta^2)$
in $f_{{\rm NL}}^{{\rm equil\,(3)}}$ is given in Appendix.
The above results reproduce those obtained for 
the theories with $G_4=0$ and $G_5=0$ \cite{nongaugali2}.

Summing up all the terms $f_{{\rm NL}}^{{\rm equil}\,(i)}$
($i=1, \cdots, 8$) and taking the largest contribution, we obtain
\begin{eqnarray}
\hspace{-0.5cm}f_{{\rm NL}}^{{\rm equil}} 
&=& \frac{85}{324} \left( 1-\frac{1}{c_s^2} \right)-\frac{10}{81} 
\frac{\lambda}{\Sigma}+\frac{55}{36}\frac{\epsilon_s}{c_s^2}
+\frac{5}{12} \frac{\eta_s}{c_s^2}-\frac{85}{54} \frac{s}{c_s^2}
+\left( \frac{20}{81} \frac{1+\lambda_{3X}}{\epsilon_s}
+\frac{65}{162 c_s^2 \epsilon_s} \right) \delta_{G3X} \nonumber \\
\hspace{-0.5cm}
& & +\left( \frac{80}{81} \frac{3+2\lambda_{4X}}{\epsilon_s}
+\frac{65}{27 c_s^2 \epsilon_s} \right) \delta_{G4XX}
+\left( \frac{20}{81 \epsilon_s} 
+\frac{65}{162 c_s^2 \epsilon_s} \right) \delta_{G5X}
+\left( \frac{20}{81} \frac{5+2\lambda_{5X}}{\epsilon_s}
+\frac{65}{162 c_s^2 \epsilon_s} \right) \delta_{G5XX}\,.
\label{fnlformula}
\end{eqnarray}
Here we have ignored the contributions such as $\delta_{G4X}$, 
$\delta_{G4\phi}$ and $\delta_{G5\phi}$ relative to 
the terms $\delta_{G4XX}/\epsilon_s$, 
$\delta_{G5X}/\epsilon_s$, and 
$\delta_{G5XX}/\epsilon_s$.

{}From Eq.~(\ref{fnlformula}) we find that 
the scalar propagation speed $c_s$
mainly determines the level of non-Gaussianities.
Expanding the term $Q$ in Eq.~(\ref{Qdef}) in terms of 
slow-variation parameters, the leading-order contribution is
\begin{eqnarray}
Q &\simeq& \Mpl^2 F [ \delta_{PX} (1+2\lambda_{PX})+6\delta_{G3X}
(1+\lambda_{3X})-2\delta_{3 \phi} 
+6\delta_{G4X}+24 \delta_{G4XX} (2+\lambda_{4X}) \nonumber \\
&&~~~~~~~~+6\delta_{G5X}+2\delta_{G5XX} (7+2\lambda_{5X})
-6\delta_{G5 \phi} ]\,,
\end{eqnarray}
where $\lambda_{iX}$ ($i=3,4,5$) are defined in Eq.~(\ref{lambdai}) and 
\begin{equation}
\lambda_{PX}=\frac{XP_{,XX}}{P_{,X}}\,.
\end{equation}
Then the scalar propagation speed squared $c_s^2=\Mpl^2 F \epsilon_s/Q$
is approximately given by 
\begin{eqnarray}
c_s^2 &\simeq& (\delta_{PX}+4\delta_{G3X}-2\delta_{G3 \phi}+6\delta_{G4X}
+20\delta_{G4XX}+4\delta_{G5X}+4\delta_{G5XX}-6\delta_{G5 \phi})/
[\delta_{PX} (1+2\lambda_{PX})+6\delta_{G3X} (1+\lambda_{3X}) 
\nonumber \\
& & -2\delta_{G3 \phi} 
+6\delta_{G4X}+24 \delta_{G4XX} (2+\lambda_{4X}) 
+6\delta_{G5X}+2\delta_{G5XX} (7+2\lambda_{5X})
-6\delta_{G5 \phi} ]\,,
\end{eqnarray}
where we have used Eq.~(\ref{eps}).
One can estimate the ratio $\lambda/\Sigma$ in 
Eq.~(\ref{fnlformula}) as follows
\begin{eqnarray}
\frac{\lambda}{\Sigma} &\simeq& 
[ \delta_{PX} (3\lambda_{PX}+2\lambda_{PXX})+3\delta_{G3X} 
(1+5\lambda_{3X}+2\lambda_{3XX})  
+6\delta_{G4XX} (9+16 \lambda_{4X}+4\lambda_{4XX})+3\delta_{G5X} 
\nonumber \\
& &+\delta_{G5XX} (27+24\lambda_{5X}+4\lambda_{5XX})] 
/3[\delta_{PX} (1+2\lambda_{PX})+6 \delta_{G3X} (1+\lambda_{3X})
-2\delta_{G 3\phi}+6\delta_{G4X} \nonumber \\
& & +24 \delta_{G4XX}
(2+\lambda_{4X})+6\delta_{G5X}-6\delta_{G5 \phi}+2\delta_{G5XX} (7+2\lambda_{5X})]\,,
\label{lamsig}
\end{eqnarray}
where 
\begin{equation}
\lambda_{PXX}=\frac{X^2P_{,XXX}}{P_{,X}},\qquad
\lambda_{3XX}=\frac{X^2 G_{3,XXX}}{G_{3,X}},\qquad
\lambda_{4XX}=\frac{X^2 G_{4,XXXX}}{G_{4,XX}},\qquad
\lambda_{5XX}=\frac{X^2G_{5,XXXX}}{G_{5,XX}}.
\end{equation}

Let us consider the case in which either of the following 
conditions is satisfied:
\begin{equation}
\lambda_{PX} \gg 1\,,\qquad 
\lambda_{3X} \gg 1\,, \qquad \lambda_{4X} \gg 1\,,\qquad
\lambda_{5X} \gg 1\,.
\end{equation}
Then one can realize $c_s^2 \ll 1$ and hence
$|f_{\rm NL}^{\rm equil}| \gg 1$.
{}From Eq.~(\ref{lamsig}) it is also 
possible to have $\lambda/\Sigma \gg 1$
if either of $\lambda_{PXX}$, $\lambda_{3XX}$,
$\lambda_{4XX}$, $\lambda_{5XX}$ is much larger than 1.

\section{Applications to concrete models of inflation}
\label{applysec}

In this section we apply our formula for the equilateral non-linear 
parameter $f_{{\rm NL}}^{{\rm equil}}$ to concrete models of 
inflation--such as 
(A) k-inflation, 
(B) k-inflation in the presence of the terms $G_i (X)$ ($i=3,4,5$),
(C) potential-driven Galileon inflation, 
(D) non-minimal coupling models, and 
(E) potential-driven inflation 
in the presence of the Gauss-Bonnet term.

\subsection{k-inflation}
\label{kinfap}

In k-inflation in which $G_3=G_4=G_5=0$ one has 
$c_s^2=1/(1+2\lambda_{PX}$) and 
$\lambda/\Sigma=(1-c_s^2)/2+2\lambda_{PXX}c_s^2/3$.
From Eq.~(\ref{fnlformula}) it follows that 
\begin{equation}
f_{{\rm NL}}^{{\rm equil}}
\simeq \frac{5}{324} \left( 1-\frac{1}{c_s^2} \right)
(17+4c_s^2)-\frac{20}{243}\frac{\lambda_{PXX}}{1+2\lambda_{PX}}
+\frac{55}{36}\frac{\epsilon_s}{c_s^2}+
\frac{5}{12} \frac{\eta_s}{c_s^2}-\frac{85}{54}\frac{s}{c_s^2}\,.
\end{equation}
Since $\epsilon=\delta_{PX}=P_{,X}X/(3\Mpl^2 H^2)$ \cite{Garriga,Hwang},
inflation occurs either around $P_{,X} \approx 0$
or $X \approx 0$.

In the former case ($P_{,X} \approx 0$), as long as
$P_{,XX}$ does not vanish at $P_{,X}=0$, we have 
$\lambda_{PX}=XP_{,XX}/P_{,X} \gg 1$ and $c_s^2 \ll 1$.
Hence large non-Gaussianities can be realized 
for the Lagrangian having a non-linear term in $X$.
The ghost condensate model $P=-X+X^2/(2M^4)$ \cite{ghost}
belongs to this class. 
For the function $P$ depending only on $X$
there is a de Sitter solution at $P_{,X}=0$.
However this is problematic because the 
scalar power spectrum ${\cal P}_{\cal R}$
diverges at the de Sitter solution 
(because $c_s=0$) \cite{kinflation}.
This problem can be avoided either by involving 
the $\phi$-dependence in $P$ (such as the 
dilatonic ghost condensate model 
$P=-X+e^{\lambda \phi/\Mpl}X^2/(2M^4)$ \cite{ghost2})
or by taking into account the terms $G_{i}(X)$ 
($i=3,4,5$) in addition to the Lagrangian $P(X)$ \cite{Galileoninf}.
We shall discuss the latter case in Sec.~\ref{kinfG}.

If inflation occurs in the region $X \approx 0$, 
whether large non-Gaussianities can be realized or not 
depends on the models. In conventional inflation driven 
by the potential energy $V(\phi)$ of the field $\phi$, 
i.e. $P=X-V(\phi)$, we have $c_s^2=1$ and 
$\lambda_{PXX}=0$, which leads to the small non-linear 
parameter $f_{\rm NL}^{\rm equil}=55 \epsilon_s/36+5\eta_s/12$.
In the DBI model where the Lagrangian 
is given by 
$P=-\sqrt{1-2f(\phi)X}/f(\phi)+1/f(\phi)-V(\phi)$ \cite{DBI},
it follows that $\lambda_{PX}=f(\phi)X/[1-2f(\phi)X]$
and $\epsilon=X/[\Mpl^2 H^2 \sqrt{1-2f(\phi)X}]$.
The non-Gaussianities can be large around the region
$2f(\phi)X \approx 1$.
If the total energy density is dominated by the 
potential energy $V$, i.e. $\Mpl^2 H^2 \approx V/3$, 
it is possible to satisfy the condition $\epsilon \ll 1$ 
for $X \ll V$ (even if $2f(\phi)X$ is close to 1).
In fact this corresponds to the regime where 
the non-linear parameter $f_{\rm NL}^{\rm equil}$ 
of the order of 10 can be achieved \cite{DBI2}. 

\subsection{k-inflation with the terms $G_i(X)$ ($i=3,4,5$)}
\label{kinfG}

In k-inflation models where the Lagrangian $P$ is a function 
of $X$ only, the problem of the de Sitter solution mentioned 
in Sec.~\ref{kinfap} can be circumvented 
by taking into account the terms $G_{i}(X)$ ($i=3,4,5$).
For the ghost condensate model $P=-X+X^2/(2M^4)$ with the term 
$G_3=\mu X/M^4$ ($\mu>0$, $M>0$) \cite{Galileoninf}, 
for example, the scalar power spectrum
is not divergent because $c_s^2 \neq 0$.
In this case, in the region where the variable
$x=X/M^4$ is close to 1, we find that $1-x \simeq \sqrt{3}\mu/\Mpl$
and $f_{\rm NL}^{\rm equil} \simeq 5/[6(1-x)] \simeq 4.62 r^{-2/3}$,
where $r \simeq 16\sqrt{6}(1-x)^{3/2}/3$.
These results match with those found in Ref.~\cite{nongaugali}
(in which detailed calculations are given).

Let us consider the following model 
\begin{equation}
P=-X+\frac{X^2}{2M^4}\,,\qquad 
G_4=\mu \frac{X^2}{M^7}\,,
\end{equation}
where $\mu$ and $M$ are positive constants.
There is a de Sitter solution characterized by the 
condition $\epsilon=\delta_{PX}+6\delta_{G4X}+12\delta_{G4XX}=0$, 
at which $P+3\Mpl^2 H^2 F-12H^2 X G_{4,X}=0$ from Eq.~(\ref{E2}).
We then obtain 
\begin{equation}
H^2=\frac{M^3}{36\mu} \frac{1-x}{x}\,,\qquad
\frac{\mu M}{\Mpl^2}=\frac{1-x}{6x^2 (3-2x)}\,,
\label{desitter}
\end{equation}
where $x=X/M^4\,(>0)$.
Provided that inflation occurs in the regime $\mu M \ll \Mpl^2$, 
the variable $x$ is close to 1 (with $x<1$). 
In the following we replace $x$
for 1 apart from the terms including $1-x$.
{}From Eq.~(\ref{desitter}) we have $\mu M/\Mpl^2 \simeq (1-x)/6$
and $H^2 \simeq M^4/(6\Mpl^2)$.
Since $Q \simeq 12\Mpl^2>0$, the no-ghost condition is satisfied.
The scalar propagation speed squared is given by 
$c_s^2 \simeq 2(1-x)/9$, so that the Laplacian instability is 
absent for $x<1$. Since $\epsilon_s \simeq 8(1-x)/3$, 
the scalar power spectrum and the tensor-to-scalar ratio are given by 
\begin{equation}
{\cal P}_{\cal R} \simeq \frac{3\sqrt{2}}{256 \pi^2}
\left( \frac{M}{\Mpl} \right)^4 
\frac{1}{(1-x)^{3/2}}\,,\qquad
r \simeq \frac{128 \sqrt{2}}{9} (1-x)^{3/2}\,.
\end{equation}
The term $\lambda/\Sigma$ in Eq.~(\ref{lamsig}) 
is estimated to be $\lambda/\Sigma \simeq 1/2$.
Since $\delta_{G4XX} \simeq (1-x)/3$, we find that 
the equilateral non-linear parameter is given by 
\begin{equation}
f_{\rm NL}^{\rm equil} \simeq 
\frac{25}{144} \frac{1}{1-x} 
\simeq 1.28 r^{-2/3}.
\label{fnlg4}
\end{equation}
For smaller $r$, $f_{\rm NL}^{\rm equil}$ gets larger.
If $r=0.01$, then $f_{\rm NL}^{\rm equil}=9.4$.
This level of non-Gaussianities can be detectable 
in future observations.

We also consider the following model
\begin{equation}
P=-X+\frac{X^2}{2M^4}\,,\qquad 
G_5=\mu \frac{X^2}{M^{10}}\,.
\end{equation}
A similar calculation shows that there is a de Sitter 
solution characterized by $\mu^2 M^4/\Mpl^6 \simeq 
27(1-x)^2/25$ and $H^2 \simeq M^4/(6\Mpl^2)$
for $x=X/M^4$ close to 1.
Since $\epsilon_s \simeq 18(1-x)/5$, $c_s^2 \simeq 3(1-x)/10$, 
$\delta_{G5X}=\delta_{G5XX} \simeq 6(1-x)/5$,  
$\lambda/\Sigma \simeq 1/2$, and $Q \simeq 12\Mpl^2>0$ 
in this case, it follows that 
\begin{equation}
f_{\rm NL}^{\rm equil} \simeq 
\frac{25}{1458} \frac{1}{1-x} 
\simeq 0.17 r^{-2/3}\,,
\end{equation}
which is about one order of magnitude smaller 
than (\ref{fnlg4}).

\subsection{Potential-driven Galileon inflation}

Let us proceed to the potential-driven inflation [$P=X-V(\phi)$] 
in the presence of the Galileon term $G_3(X) \propto X^n$ ($n>0$) \cite{Elliston}.
The inflationary dynamics for the case $n=1$
were studied in Refs.~\cite{Kamada,Elliston}.
Since $\lambda_{PX}=0$ and $\lambda_{3X}=n-1$ we have
\begin{equation}
c_s^2 =\frac{\delta_{PX}+4\delta_{G3X}}
{\delta_{PX}+6n\delta_{G3X}}\,.
\end{equation}
In the regime $\delta_{G3X} \gg \delta_{PX}$
it follows that $c_s^2 \simeq 2/(3n)$ and 
hence $c_s^2 \ll 1$ for $n \gg 1$.
In this case one has $\epsilon_s \simeq 4\delta_{G3X}$
and $\lambda/\Sigma \simeq n/3$, so that the non-linear
parameter (\ref{fnlformula}) is given by 
\begin{equation}
f_{\rm NL}^{\rm equil} \simeq -\frac{865}{3888}n\,.
\end{equation}

In the presence of the term $G_4(X) \propto X^n$ 
the slow-variation parameter $\delta_{G4XX}$
is related with $\delta_{G4X}$, via
$\delta_{G4XX}=(n-1)\delta_{G4X}$.
If $n \gg 1$, 
one has $c_s^2 \simeq 5/(6n)$, 
$\lambda/\Sigma \simeq n/3$, 
$\epsilon_s \simeq 20\delta_{G4XX}$ 
in the regime $1 \gg \delta_{G4XX} \gg \delta_{PX}$, and 
hence $f_{\rm NL}^{\rm equil} \simeq -137n/1215$.
Similarly, in the case where 
$G_5(X) \propto X^n$ with $n \gg 1$, 
it follows that $f_{\rm NL}^{\rm equil} \simeq 
-155n/1944$ for $1 \gg \delta_{G5XX} \gg \delta_{PX}$.

\subsection{Nonminimal coupling models}

The nonminimal coupling of the scalar field $\phi$
with the Ricci scalar $R$ corresponds to the choice
$G_4=F(\phi)$, where $F(\phi)$ is an arbitrary 
function in terms of $\phi$.
In the absence of the terms $G_3$ and $G_5$, 
using $\delta_{G4X}=\delta_{G4XX}=0$, the 
scalar propagation speed squared is $c_s^2=1/(1+2\lambda_{PX})$.
Hence, as in the case of k-inflation, we require 
$\lambda_{PX} \gg 1$ to realize large non-Gaussianities.
For the theories in which $P$ does not have non-linear 
terms in $X$ we have $c_s^2=1$ and
\begin{equation}
f_{\rm NL}^{\rm equil}={\cal O}(\epsilon_s, \eta_s)\,.
\label{fNLs}
\end{equation}
For example, Brans-Dicke theories described by the action 
$P=\omega_{\rm BD} \Mpl X/\phi-V(\phi)$ \cite{BD} belong 
to this class. Hence the non-Gaussianities in those theories 
are small (including $f(R)$ gravity where
the Brans-Dicke parameter $\omega_{\rm BD}$ 
is 0 \cite{Ohanlon}).

The models with field derivative couplings 
recently studied in Refs.~\cite{Germani,Germani2}
correspond to $G_{5}=F(\phi)$, in which case 
$\delta_{G5X}=0$ and $\delta_{G5XX}=0$.
In the absence of the terms $G_3$ and $G_4$ we have 
\begin{equation}
c_s^2=\frac{\delta_{PX}-6\delta_{G5\phi}}
{\delta_{PX}(1+2\lambda_{PX})-6\delta_{G5\phi}}.
\end{equation}
If the Lagrangian does not include non-linear
terms in $X$ (like the models discussed in 
Ref.~\cite{Germani}), then $c_s^2=1$ and 
$f_{\rm NL}^{\rm equil}={\cal O}(\epsilon_s, \eta_s)$.
This is consistent with the results 
found in Ref.~\cite{Germani2}.

\subsection{Potential-driven inflation with a Gauss-Bonnet term}

The action (\ref{action}) even covers the Gauss-Bonnet coupling 
of the form $-\xi(\phi){\cal G}$, where
${\cal G}=R^{2}-4R_{\alpha\beta}R^{\alpha\beta}
+R_{\alpha\beta\gamma\delta}R^{\alpha\beta\gamma\delta}$
is the Gauss-Bonnet term.
If one chooses the following combination of $P$, $G_{3}$, $G_{4}$,
and $G_{5}$ 
\begin{eqnarray}
& & P=-8\xi^{(4)} (\phi) X^2 (3-\ln X)\,,\qquad
G_3=-4\xi^{(3)} (\phi) X (7-3\ln X)\,,\label{GBcom0} \nonumber \\
& & G_4=-4\xi^{(2)} (\phi) X (2-\ln X)\,,\qquad
G_5=4\xi^{(1)} (\phi) \ln X\,,
\label{GBcom}
\end{eqnarray}
where $\xi^{(n)}(\phi)=\partial^n \xi (\phi)/\partial \phi^n$,
one can show that the field equations following from 
this Lagrangian are equivalent to those derived by the Lagrangian 
$-\xi(\phi){\cal G}$ \cite{KYY}.

Let us consider the case of potential-driven inflation in which
the Lagrangian $X-V(\phi)$ is added to $P$ in Eq.~(\ref{GBcom}).
Using the choice of the functions in Eq.~(\ref{GBcom}), 
we find that the scalar propagation speed squared 
is given by 
\begin{equation}
c_{s}^{2}=1-\frac{64\delta_{\xi}^{2}}{\delta_{X}}\,
(6\delta_{\xi}+\delta_{X})\,,
\end{equation}
where $\delta_X=X/(\Mpl^2 H^2)$ and
$\delta_{\xi}=H\dot{\xi}/\Mpl^2$.
This expression agrees with the one 
found in Ref.~\cite{Elliston}. 
Since $c_s^2-1$ is a second-order quantity, the 
scalar propagation speed is very close to 1.
Furthermore we can show that, in this case, $\lambda/\Sigma=0$,
and that the terms in Eq.~(\ref{fnlformula})
coming from the functions $\delta_{G3X}$,
$\delta_{G4XX}$, $\delta_{G5X}$, and $\delta_{G5XX}$
are of the order of $\epsilon$, namely proportional to $\delta_{\xi}\,(4\epsilon+2\eta_{\xi}-\eta_{X})/\delta_{X}$.
Using $\epsilon_s \simeq \delta_X$, the leading contribution 
to $f_{\rm NL}^{\rm equil}$ is given by 
\begin{equation}
f_{\rm NL}^{\rm equil}=\frac{55}{36}\delta_X
+\frac{5}{12}\eta_X+\frac{275}{81}
\frac{\delta_{\xi}}{\delta_{X}}
(4\epsilon+2\eta_{\xi}-\eta_{X})\,,
\end{equation}
where $\eta_{\xi}=\dot{\delta}_{\xi}/(H \delta_{\xi})$ 
and $\eta_X=\dot{\delta}_{X}/(H\delta_X)$.
Hence the non-linear parameter in this model is small.

\section{Conclusions}
\label{concludesec}

In this paper we have evaluated the primordial 
non-Gaussianities generated during inflation 
for the models described by 
the action (\ref{action}).
Our analysis is general enough in that it covers 
a wide variety of single-field models having second-order 
equations of motion.

The procedure to obtain the three-point correlation 
function of curvature perturbations is similar to 
that given in Ref.~\cite{nongaugali2}.
There are 8 shape functions which contribute 
to the scalar non-Gaussianities.
Among them, five functions are already known to give
rise to the equilateral type of non-Gaussianities. 
We studied the shapes of the remaining three functions and 
found that they can be well approximated by 
the same type as well. 
Hence the dominant contribution to the three-point 
correlation function comes from the equilateral configuration.

We derived the equilateral non-linear parameter
$f_{\rm NL}^{\rm equil}$ under the approximation that the 
slow-variation parameters defined in Eq.~(\ref{slowva})
are much smaller than 1.
The formula (\ref{fnlformula}) is valid for any 
quasi de Sitter background and thus it is convenient 
to apply it to concrete single-field models of inflation.
In fact we used this formula to a number of models such as 
(A) k-inflation, 
(B) k-inflation with the terms $G_i(X)$ ($i=3,4,5$),
(C) potential-driven Galileon inflation,
(D) nonminimal coupling models without a non-linear term in $X$, 
(E) potential-driven inflation with a Gauss-Bonnet term.
In the models (D) and (E) we showed that 
$c_s^2$ is close to 1 and hence 
the non-Gaussianities are small.
However, for the models (A), (B), (C), it is possible to 
realize $|f_{\rm NL}^{\rm equil}| \gg 1$ depending 
on the choice of the functions $P, G_i$ ($i=3,4,5$).

The potential detectability of non-Gaussianities 
in future observations will allow us to distinguish 
between different inflationary models.
In particular we expect that the precise measurement of 
$f_{\rm NL}^{\rm equil}$
as well as $n_{\cal R}$ and $r$ 
will provide significant implications for the viability 
of single-field models in which $c_s^2$ is much smaller than 1.

%%%%%%%%%%%%%%%%%%%%%%%%%%%%%%%%%%%%%
\section*{ACKNOWLEDGEMENTS}
\label{acknow} The work of A.\,D.\,F.\ and S.\,T.\ was supported
by the Grant-in-Aid for Scientific Research Fund of the JSPS Nos.~10271
and 30318802. S.\,T.\ also thanks financial support for the Grant-in-Aid
for Scientific Research on Innovative Areas (No.~21111006). 
We thank the members of the Institute for Fundamental Study 
for warm hospitality during our stay in Naresuan University.
%%%%%%%%%%%%%%%%%%%%%%%%%%%%%%%%%%%%%

\vspace{0.5cm}
{\it Note added}--While we were completing this work, we became aware of 
the paper by Gao and Steer \cite{Gaonon} who calculated primordial 
non-Gaussianities in the same model as ours. Their formula for the bispectrum (98)
is consistent with our formula (\ref{AcalR}).

\appendix

%%%%%%%%%%%%%%%%%%%%%%%%%%
\section{The second-order term 
in $f_{\rm NL}^{\rm equil\,(3)}$}\label{appen}
%%%%%%%%%%%%%%%%%%%%%%%%%%

The second-order term $\xi(\delta^2)$ in Eq.~(\ref{eq:fNL3})
is given by 
\begin{eqnarray}
\xi (\delta^2) &=&
[(6+4\lambda_{3X})\delta_{G4X}+8(3+\lambda_{3X}
+\lambda_{4X})\delta_{G4XX}+3(2+\lambda_{3X})\delta_{G5X}
+(9+2\lambda_{3X}+2\lambda_{5X})\delta_{G5XX}
-\delta_{G4\phi} (1+\lambda_{3X})\nonumber \\
& &
-2\delta_{G5\phi}
(3+2\lambda_{3X})]\delta_{G3X}
+8\delta_{G4X}^2+[16 (5+2\lambda_{4X})\delta_{G4XX}
+18\delta_{G5X}+8(4+\lambda_{5X})\delta_{G5XX} 
-2\delta_{G4\phi}\nonumber \\
& &-16\delta_{G5\phi}]\delta_{G4X}
+64(2+\lambda_{4X})\delta_{G4XX}^2+[24(3+\lambda_{4X})
\delta_{G5X}+8(11+2\lambda_{4X}+2\lambda_{5X})\delta_{G5XX}
-8\delta_{G4\phi} (2+\lambda_{4X}) \nonumber \\
& &-16 \delta_{G5\phi}(5+2\lambda_{4X})]\delta_{G4XX}
+9\delta_{G5X}^2+[3(9+2\lambda_{5X})\delta_{G5XX}
-3\delta_{G4\phi}-18\delta_{G5\phi}]\delta_{G5X}
+2(7+2\lambda_{5X})\delta_{G5XX}^2 \nonumber \\
& & -[(7+2\lambda_{5X})\delta_{G4\phi}
+8(4+\lambda_{5X})\delta_{G5\phi}]\delta_{G5XX}+
8\delta_{G5\phi}^2+2\delta_{G4\phi}\delta_{G5\phi}\,.
\end{eqnarray}

\end{document}